\documentclass[
preprint,
 amsmath,amssymb,
prfluids,
floatfix,
]{revtex4-2}

\usepackage{graphicx}
\usepackage{dcolumn}
\usepackage{bm}

\usepackage[caption=false]{subfig}
\usepackage{multirow}
\usepackage{color}

\begin{document}


\title{Life-cycle of streaks in the buffer layer of wall-bounded turbulence}

\author{H. Jane Bae}
 \email{jbae@caltech.edu}
 \affiliation{School of Engineering and Applied Sciences, Harvard University, Cambridge, MA 02139, USA}
 \affiliation{Graduate Aerospace Laboratories, California Institute of Technology, Pasadena, CA 91125, USA}
\author{Myoungkyu Lee}%
 \email{mnlee@sandia.gov}
 \affiliation{Combustion Research Facility, Sandia National Laboratories, Livermore, CA 94550, USA}

\date{\today}

\begin{abstract}
Streaks in the buffer layer of wall-bounded turbulence are tracked in time to study their life-cycle. Spatially and temporally resolved direct numerical simulation data is used to analyze the strong wall-parallel movements conditioned to low-speed streamwise flow. The analysis of the streaks shows that there is a clear distinction between wall-attached and detached streaks, and that the former can be further categorized into streaks that are contained in the buffer layer and the ones that reach the outer region. The results reveal that streaks are born in the buffer layer, coalescing with each other to create larger streaks that are still attached to the wall. Once the streak becomes large enough, it starts to meander due to the large streamwise-to-wall-normal aspect ratio, and consequently the elongation in the streamwise direction, which makes it more difficult for the streak to be oriented strictly in the streamwise direction. While the continuous interaction of the streaks allows the super-structure to span extremely long temporal and length scales, individual streak components are relatively small and short-lived. Tall-attached streaks eventually split into wall-attached and wall-detached components. These wall-detached streaks have a strong wall-normal velocity away from the wall, similar to ejections or bursts observed in the literature. Conditionally averaging the flow fields to these split events show that the detached streak has not only a larger wall-normal velocity compared to the wall-attached counterpart, it also has a larger (less negative) streamwise velocity, similar to the velocity field at the tip of a vortex cluster. 
\end{abstract}

\maketitle

\section{Introduction}

Among the many organized structures observed in near-wall turbulent flows, streaks, defined as regions of slowly moving fluid elongated in the direction of the mean flow, are considered to be of major importance for their role in the regeneration of turbulent energy. In the flow-visualization study by Kline \emph{et al.}~\cite{Kline1967}, it was established that the important characteristics of the near-wall region of wall-bounded turbulent flows are the streak formation in the viscous sublayer and the subsequent ejection of the low-velocity fluid to the outer region of the flow. 

Flow visualization employing dye, particles, bubbles, and smoke has played a major role in the study of turbulent coherent motions. Quantitative analyses of flow-visualization studies by Corino \& Brodkey~\cite{Corino1969}, Kim \emph{et al.}~\cite{Kim1971}, and Grass~\cite{Grass1971} indicated that the ejection of low-velocity fluid from the wall region was associated with a major part of Reynolds stress and turbulent energy production. According to these studies, the low-velocity streak slowly lifts away from the wall, at which time, the streak filament begins to oscillate in both the spanwise and wall-normal directions. The bursting process continues as the loops of the streak filaments eject away from the wall. Finally, the ejected streak filaments eventually break up in a chaotic process. 

Since most of the turbulence production in the near-wall region occurs during those bursts, many following studies have been performed using probes to measure the velocity and pressure fields associated with bursts. Examples of works using conditional sampling techniques to identify the structures involved in the bursting process using probe measurements can be found in the $uv$-quadrant method \cite{Wallace1972,Willmarth1972}, $u$-level detection method \cite{Lu1973}, the VITA (variable-interval time average) method \cite{Blackwelder1976}, and the VISA (variable-interval space average) method \cite{Kim1985b}. Key aspects of the conditional sampling methods for turbulence structures are reviewed by Bogard \& Tiederman~\cite{Bogard1986}, among others.

The advent of particle-image velocimetry (PIV) provided two-dimensional instantaneous flow fields, which allowed a more in-depth analysis of instantaneous flow fields compared to probe measurements. PIV experiments led to studies linking groups of ejections to low-momentum streaks \cite{Adrian1991,Adrian2005}. Tomkins \& Adrian~\cite{Tomkins2003} and Ganapathisubramani \emph{et al.}~\cite{Ganapathisubramani2003} used PIV data on wall-parallel planes in a turbulent boundary layer to quantify the contribution to the Reynolds stresses by packets of hairpins. Kevin \emph{et al.}~\cite{Kevin2017,Kevin2019} used PIV measurements from spanwise homogeneous and heterogeneous boundary layers to identify streak meandering. Three-dimensional measurements using high-speed PIV coupled with Taylor’s hypothesis \cite{Dennis2011a,Dennis2011b} has been utilized to track vortices and long structures in turbulent boundary layers. 

Simultaneously, the rise of computational power has allowed access to fully resolved three-dimensional data sets, which led to the study of instantaneous three-dimensional coherent structures extracted from direct numerical simulations (DNS) of wall-bounded flow \cite{Robinson1991}, such as the characterization of clusters of vortices \cite{Jeong1997,Moisy2004,Tanahashi2004,delAlamo2006}, the generalized three-dimensional quadrant analysis \cite{Lozano-Duran2012}, and the study of intense regions of individual velocity components \cite{delAlamo2003,Abe2004,Sillero2014}. Temporally resolved data sets allow an additional dimension in the coherent structure analysis, providing a full picture of how the structures evolve in space and time \cite{Toh2005,Lozano-Duran2014,Abe2018} and how information transfers between different structures \cite{Lozano-Duran2020}. 

These observations led to structural models that explain the dynamics of wall-bounded turbulence. The most established models describe motions in the buffer layer; examples include the papers by Jim\'enez \& Moin~\cite{Jimenez1991}, Jim\'enez \& Pinelli~\cite{Jimenez1999}, Schoppa \& Hussain~\cite{Schoppa2002}, and Kawahara \emph{et al.}~\cite{Kawahara2012}, among others. While a lot of effort has been devoted to studying the dynamics of the buffer layer, most of them have been in idealized conditions with simplified dynamics, with most focus given in identifying the dynamics of a single isolated streak. Furthermore, studies of coherent structures mentioned throughout this introduction have mainly focused on vortical structures and Reynolds stress, and with good reason. Reynolds stress is strongly related to the turbulent kinetic energy production -- the understanding of the mechanism of Reynolds stress generation is central to predicting the effects of turbulence in a wide variety of natural settings and engineering applications. Quadrant analysis classifies the Reynolds shear stress into four categories based on the sign of the streamwise ($u$) and wall-normal fluctuations ($v$). Q2 ($u<0$, $v>0$) and Q4 ($u>0$, $v<0$) events play important roles in most of the structural models explaining how turbulent kinetic energy and momentum are redistributed. These models are loosely based on the attached-eddy hypothesis by Townsend~\cite{Townsend1961} and involve wall-attached vortical loops growing from the wall into the outer region \cite{Perry1986}. The study of these structures led to further understanding of near-wall turbulence, especially with the characterization in space and time \cite{Lozano-Duran2014}, but research focused on the kinematics and the dynamics of the temporally and spatially resolved interaction of the streaks in the buffer layer is still incomplete. A broader view of the full life-cycle of streaks, involving ejections and bursts, could complement the ongoing study of Reynolds stresses and vortical clusters.

The goal of this paper is to study the life-cycle of streaks, classified in terms of the streamwise and spanwise fluctuations, and to study the time evolution of the size and meandering of the streaks. We classify the streaks into tall-attached, detached, and buffer layer structures and study how the detached streaks correlate with the Q2 or bursting events. For this purpose, we track streaks in spatially and temporally resolved flow fields of low-Reynolds-number turbulent channel flow. The remainder of this paper is organized as follows. The details of the numerical simulations are given in \S\ref{sec:numerics}. The methodology used to identify and track individual streak structures are introduced in \S\ref{sec:streak}. Static and temporal analysis of streaks are given in \S\ref{sec:results:static} and \S\ref{sec:results:temp}, respectively. Finally, the summary of the work and the conclusions are offered in \S\ref{sec:conclusions}.

\section{Numerical experiment}\label{sec:numerics}
 
A DNS of a channel flow at friction Reynolds number $Re_\tau=u_\tau \delta/\nu\approx 186$ is performed, where $\nu$ is the kinematic viscosity, $\delta$ is the channel half-height, and $u_\tau$ is the friction velocity at the wall.
As the small-scale motions near the wall is universal \cite{Jimenez1991,Jimenez1999,Hwang2013,Lee2019}, this Reynolds number is ideal for studying the buffer layer streaks without any interference from the outer-region large-scale motions.
Throughout the paper, $x$, $y$, and $z$ denote the streamwise, wall-normal, and spanwise directions, respectively. The corresponding fluctuating velocity components are $u$, $v$, and $w$. The root-mean-squared (r.m.s.) intensities are given by $u'$, $v'$, and $w'$, respectively. The only non-zero mean velocity is in the streamwise component, denoted $U(y)$. The simulations are computed with a staggered second-order finite difference~\cite{Orlandi2000} and a fractional-step method~\cite{Kim1985a} with a third-order Runge-Kutta time-advancing scheme~\cite{Wray1990}. Periodic boundary conditions are imposed in the streamwise and spanwise directions and the no-slip and no-penetration boundary conditions are used at the top and bottom walls. The code has been validated in previous studies in turbulent channel flows~\cite{Bae2018,Bae2019,Lozano-Duran2019a,Lozano-Duran2019b} and flat-plate boundary layers \cite{Lozano-Duran2018}.

The numerical domain is $8\pi\delta \times 2\delta\times 3\pi\delta$ in the streamwise, wall-normal, and spanwise directions, respectively. While this domain is not long enough to capture the longest of near-wall streaks that are of the order of $10^{4-5}$ wall units long, these streaks are known to be formed by coalescence of several shorter ones \cite{Jimenez2004}, and thus for our purposes, the current domain is adequate for tracking individual streak structures. The domain is discretized using $768$, $130$, and $288$ grid points in the $x$, $y$ and $z$ directions, respectively. This corresponds to uniform grid spacings in the streamwise and spanwise directions of $\Delta x^+ = 6$, $\Delta z^+ = 3.5$ and a wall-normal grid stretched away from the wall using a hyperbolic tangent with $\min(\Delta y^+) = 0.16$ and $\max(\Delta y^+) = 7.2$, where the superscript $+$ denotes wall units defined in terms of $\nu$ and $u_\tau$. The simulations were run for 100 eddy turnover times (defined as $\delta/u_\tau$) after transients to compute the mean quantities. The analysis of the temporal evolution of the flow requires storing approximately $2\times10^3$ snapshots spaced $\Delta t^+\approx 1$ apart. The time-resolved data set covers approximately $11.4 \delta/u_\tau$, whereas the longest lifetimes Reynolds stress structures are less than $2 \delta/u_\tau$ \cite{Lozano-Duran2014}. While individual streaks have longer lifetimes than those of the Reynolds stress structures, our analysis in \S~\ref{sec:results:temp} shows that the current temporal range is enough to capture the longest life-cycle of streaks as well. 

\section{Streak identification and tracking}\label{sec:streak}

\begin{figure}
\centering
\subfloat[]{\includegraphics[height=0.38\textwidth]{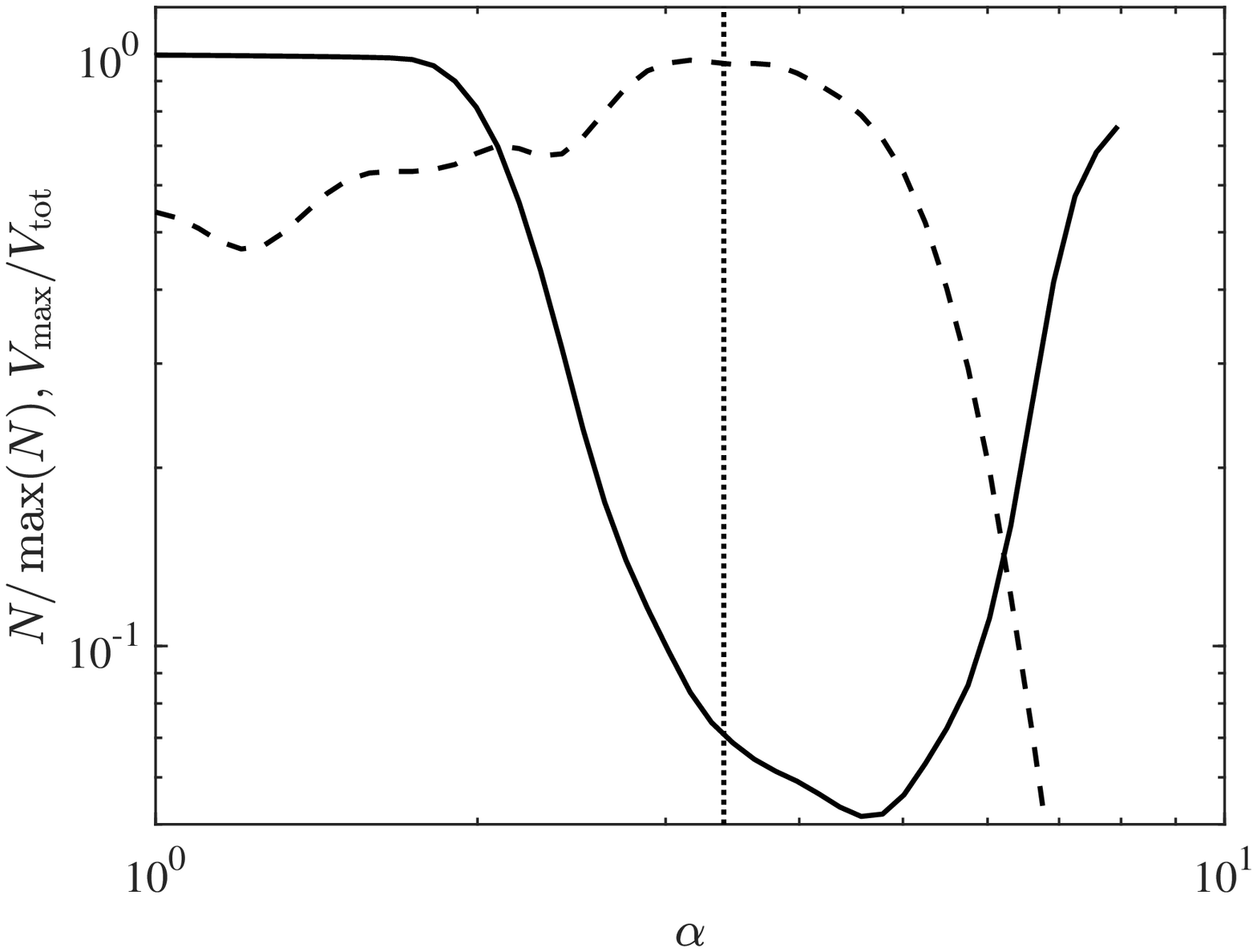}}
\hspace{0.2cm}
\subfloat[]{\includegraphics[height=0.38\textwidth]{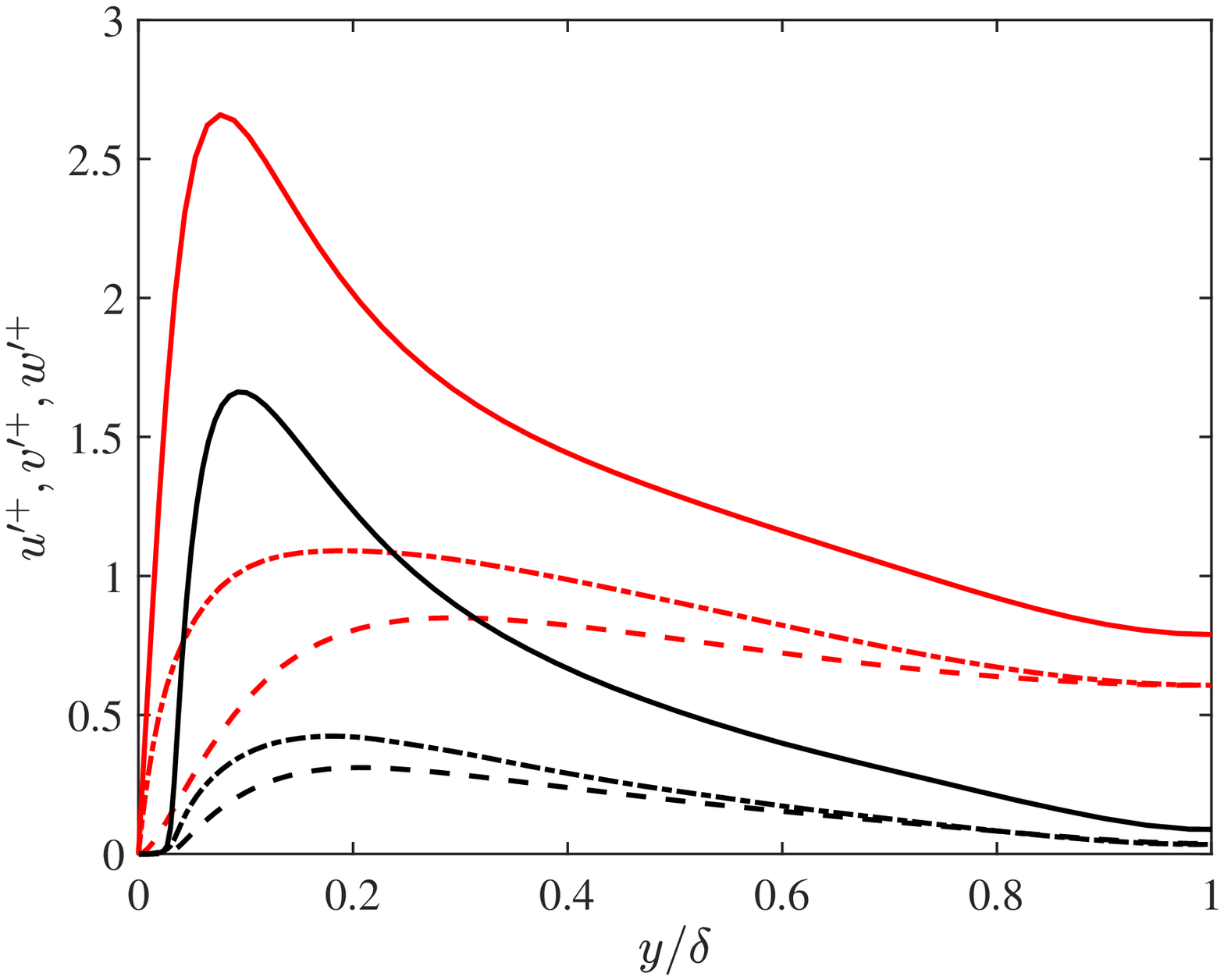}}\\
\subfloat[]{\includegraphics[width=0.60\textwidth]{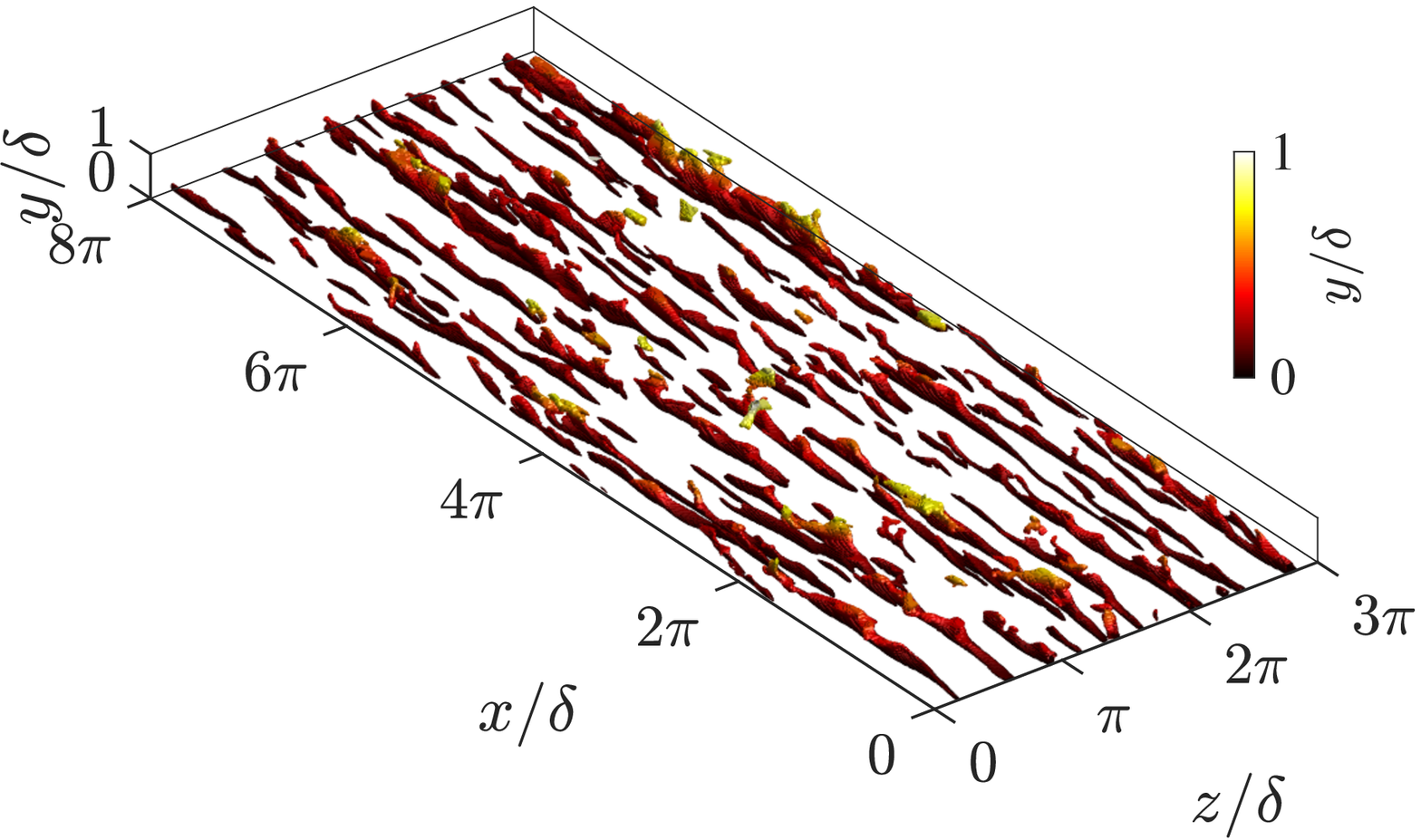}}
\caption{({a}) Percolation diagram for the identification of streaks. Curves indicate ratio of the volume of the largest object to the volume of all identified objects, $V_{\max}/V_{\text{tot}}$ (solid line) and ratio of the number of identified objects to the maximum number of objects, $N/\max(N)$, (dashed line). The vertical dotted line indicates the chosen threshold, $\alpha=3.4$. ({b}) Streamwise (solid), spanwise (dot-dashed) and wall-normal (dashed) r.m.s. intensity contribution from the full channel (red) and only from the identified streaks (black).  ({c}) Streaks identified in a single snapshot using threshold $\alpha = 3.4$ colored by the distance to the wall.}
\label{fig:percolation}
\end{figure}   
In the present work, we identify streaks as a connected set of points within the full domain where the streamwise velocity fluctuation is less than zero and the wall-parallel velocity magnitude exceeds a given threshold, i.e.,
\begin{equation}
    \left\{(x,y,z): u(x,y,z) < 0\ \text{and}\ \sqrt{u^2(x,y,z)+w^2(x,y,z)} > \alpha u_\tau \right\},
    \label{eq:streak_def}
\end{equation}
where $\alpha$ is a threshold value. This is similar to the traditional definition of streaks, but with additional contribution from the spanwise fluctuations. The spanwise fluctuations are included to account for the strong meandering events of streaks, where the spanwise velocity component would become dominant. Disregarding the spanwise component, in this case, would result in these meandering events being classified as streaks separating from one another, which is undesirable. Connectivity in space is defined in terms of six orthogonal neighbors in the Cartesian mesh of the DNS. 

The threshold $\alpha$ can be obtained from a percolation analysis~\cite{Moisy2004,delAlamo2006,Lozano-Duran2012} (see Figure~\ref{fig:percolation}{a}). Percolation theory describes the behavior of a network when nodes or links are removed. Here, percolation analysis is applied to the variations of the volume of the connected objects extracted by Eq. \eqref{eq:streak_def} with the threshold parameter $\alpha$. When $\alpha$ is very large, the identification only yields a few small streaks. Decreasing $\alpha$ introduces new streaks while the previously identified ones grow in size. At first, the ratio of the volume of the largest streak to the volume of all identified streaks, $V_{\text{max}}/V_{\text{tot}}$ decreases, but it increases rapidly as streaks merge and form larger streaks. This percolation crisis occurs around $1\lesssim\alpha\lesssim4$. The value of $\alpha=3.4$, which lies within these bounds was chosen to maximize the number of streaks identified ($N$). Small changes to the parameter $\alpha$ do not change the conclusions of the current study (not shown), and thus we only focus on results given by $\alpha=3.4$. This threshold corresponds to about $1.4$ times the peak streamwise turbulence intensity for the $Re_\tau=186$ case. This ensures that the majority of the streaks identified are located in the buffer layer (typically defined as $5\le y^+ \le 30\sim100$), with only very strong events identified in the outer region, which is in line with our study of buffer layer streaks and their connection to these strong outer layer streaks.

Once the streaks satisfying the threshold are identified, streaks with volume less than $30^3$ wall units are discarded to reduce noise in the identification of streak-to-streak interactions. Figure~\ref{fig:percolation}(b) shows the contribution of the turbulent intensities from the streaks compared to the full channel for each velocity component. The streaks carry more than 60\% of the streamwise turbulence intensity in the peak of the buffer region. Also, the streaks are responsible for 20\% of the total kinetic energy of the entire domain while only taking up 2\% of the volume. The identified streaks for a single snapshot are shown in Figure~\ref{fig:percolation}({c}), which shows that indeed the majority of the streaks lie in the buffer region, with some streaks identified in the outer region.

\begin{figure}
\centering
\subfloat[]{\includegraphics[width=0.6\textwidth]{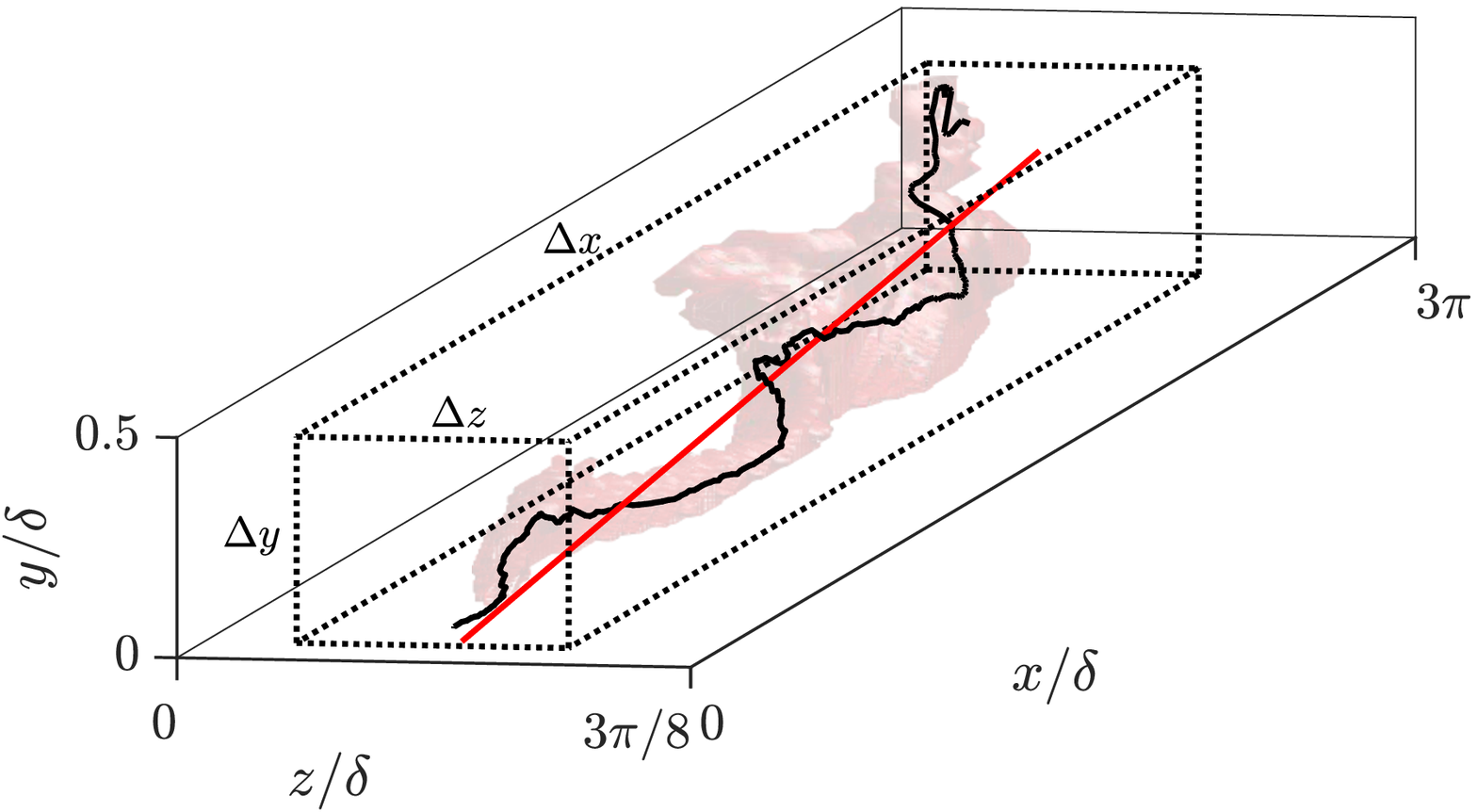}}\\
\subfloat[]{\includegraphics[width=0.6\textwidth]{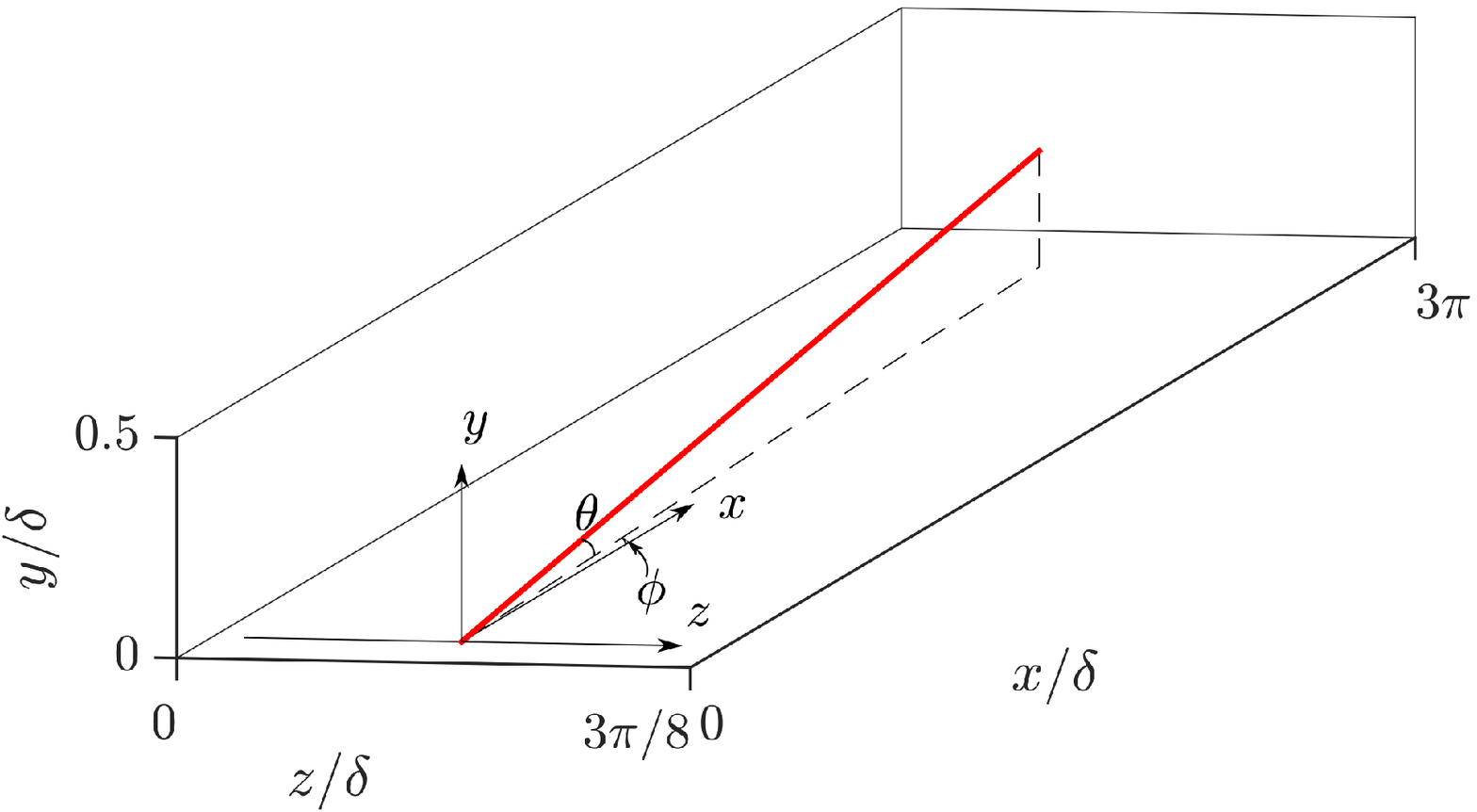}}
\caption{({a}) Spine (black solid line) and its linear fit (red solid line) for a single identified streak. Dotted lines indicate the bounding box of the streak. ({b}) Diagram depicting the azimuth ($\phi$) and elevation ($\theta$) angles of the linear fit for the case shown in ({a}).}
\label{fig:spine}
\end{figure}

For each streak identified, a few key quantities are computed to characterize the streak, namely size, volume, meandering, and orientation of the streak. First, the size of the streak is given by the dimensions of the bounding box ($\Delta x\times\Delta y\times\Delta z$), which is defined as the smallest box that can encapsulate the streak (Figure~\ref{fig:spine}{a}). The volume of the streak is computed as the volume the streak occupies in the domain, not by the volume of the bounding box. The spine of the streak, $\Sigma = \left\{(x_s,y_s,z_s)\right\}$, is computed as the geometric center in the $yz$-plane for each streamwise location the streak occupies. The spine is then fitted into a line $\Lambda_s = \left\{(x,y,z): (x-\bar{x}_s)/a_x = (y-\bar{y}_s)/a_y = (z-\bar{z}_s)/a_z\right\}$ that minimizes the $L_2$-norm, as shown in Figure~\ref{fig:spine}({a}). Here, ($\bar{x}_s$, $\bar{y}_s$, $\bar{z}_s$) is the mean $x$, $y$, $z$ coordinate of the spine, and $(a_x, a_y, a_z)$ is the direction vector of the line $\Lambda_s$. The meandering of the streak is quantified as the deviation of the spine to the linear fit, i.e., the mean-squared error defined as 
\begin{equation}
\|\mathcal{E}\| \equiv \frac{\int_{\vec{\boldsymbol{x}}_s\in\Sigma} \min_{\vec{\boldsymbol{x}}\in \Lambda_s} \left(\vec{\boldsymbol{x}}_s-\vec{\boldsymbol{x}}\right)^{2} \mathrm{d}S }{\Delta x_s\Delta y_s \Delta z_s},
\end{equation}
where $\vec{\boldsymbol{x}} = (x,y,z)$, $\vec{\boldsymbol{x}}_s = (x_s,y_s,z_s)$, and $\Delta x_s$, $\Delta y_s$, and $\Delta z_s$ is the streamwise, wall-normal, and spanwise length of the spine, respectively \cite{Kevin2019}. This definition of meandering ensures that resizing of the streaks (multiplication by a constant factor in all dimensions) would not affect the meandering coefficient $\|\mathcal{E}\|$. Using the spine and its linear fit to quantify meandering ensures that the width of the streak and the linear orientation of the spines does not play a role. The orientation of the streak is quantified by the azimuth and elevation angles (see Figure~\ref{fig:spine}{b}), which are computed as $\phi = \tan^{-1}(a_z/a_x)$ and $\theta = \tan^{-1}(a_y/a_x)$, respectively. The azimuth angle indicates how much the streak deviates in the spanwise direction whereas the elevation indicates how much it deviates in the wall-normal direction. 

\begin{figure}
\centering
\includegraphics[width=0.9\textwidth]{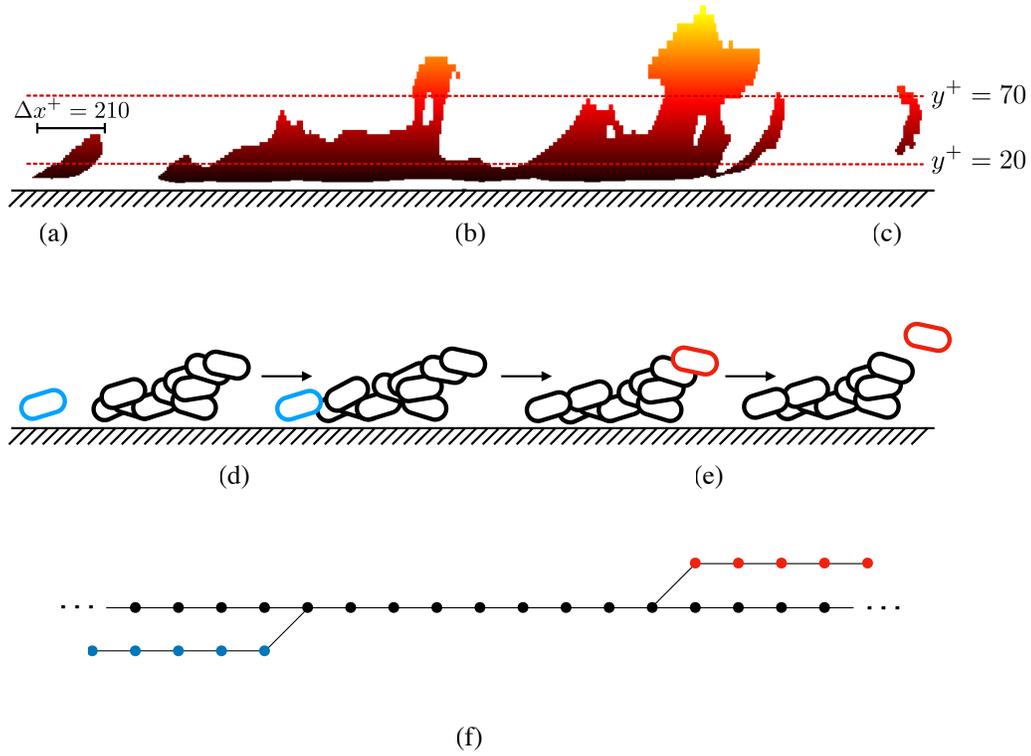}
\caption{Examples of (a) buffer-layer, (b) tall-attached, and (c) detached streaks. Sketch of a (d) merger and (e) split event. (f) Graph associated with the evolution shown in (d,e).}
\label{fig:diagram}
\end{figure}

For temporal analysis, we track the evolution of individual streaks following the method in Lozano-Dur\'an \& Jim\'enez \cite{Lozano-Duran2014}. The streaks from every two consecutive snapshots are evaluated for overlaps between them. When comparing consecutive snapshots for overlap, the preceding flow field is shifted by $U(y)\Delta t$ in the streamwise direction to account for the advection, which is mostly due to the mean flow \cite{Taylor1938,Kim1993,Krogstad1998}. All of the streaks with nonzero overlap are considered connected. The connections in time identified are then organized into graphs such that each streak (at all time steps) is considered a vertex and the connection between consecutive times is considered an edge. This way, the temporal evolution of a single streak can be identified as a connected component within this graph. If the streak has more than one backward connection, i.e. more than one streak is connected to a single streak in the next time step, it is considered a merger event. Conversely, if a streak has more than one forward connection, it is considered a split event. See Figure~\ref{fig:diagram}(d--f) for a sketch of merging and splitting events and the corresponding graph. A single connected component of the graph signifies the evolution of all of the structures that interact with each other at some point in time and may have multiple merger and split events, possibly making the structure more complex. For simplicity, each of these connected components will be simply referred to as a `graph', and the superset graph containing all the connected components will be referred to as the `supergraph'.  

\begin{figure}
\centering
\includegraphics[width=0.9\textwidth]{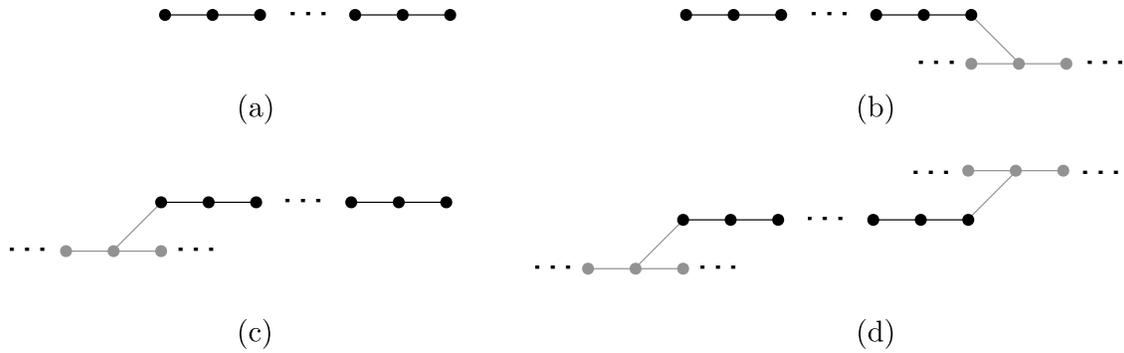}
\caption{Examples of (a) primary, (b) incoming, (c) outgoing, and (d) connector branches.}
\label{fig:branches}
\end{figure}  

Graphs are then organized into `branches' representing individual structures within the graph. For each merger and split event, i.e., when a node has more than one forward or backward edge, the weight $\Delta V/V_o$ of each of the edges is computed, where $\Delta V$ is the absolute value of the volume difference in the two streaks in the two end nodes of the edge and $V_o$ is the volume overlap. The edge with the smallest weight is considered the primary connection and the rest are considered secondary. The secondary connections in a merger event are considered the end of that branch that merged into a larger branch, and the secondary connections in a split event are newly created branches that split from the main branch. This way, the complex spatio-temporal interaction between various streaks can be broken down into individual streaks. Note that a merger or split event is a single event in time and, thus, identified as a node in the graph whereas a branch is a series of connected nodes and identify the evolution of a streak in time.  

Each branch is classified as `primary', `incoming', `outgoing', and `connector' depending on how they are created and destroyed (see Figure~\ref{fig:branches}). Primary branches have no forward or backward connections and are created from and dissipate into the turbulent background. Outgoing and incoming branches either split from or merge into another streak, respectively. Connector branches start and end in another streak, created through a split event and dying in a merger event, effectively connecting two streaks. See Lozano-Dur\'an \emph{et al.}~\cite{Lozano-Duran2014} for a more comprehensive explanation of the tracking algorithm.

For the current data set, graphs that traverse the entirety of the temporal length ($T^+ \approx 2050$, $Tu_\tau/\delta\approx 11.4$) account for only two out of the 5,810 identified graphs. However, due to the complexity of these graphs, the two graphs account for 73.9\% of all streak structures. The branches that compose these graphs do not span a long time period, with a mean duration of 28 wall units. This indicates that the extremely long-lasting graph structures are sustained from the continuous merging of individual branches. This is also evident in the breakdown of branch categories. The primary branches account for 9.2\% of all the branches, composing only a small fraction of all branches. Incoming, outgoing, and connector branches account for 14.9, 37.2, and 38.7\% of the branches, respectively.

\section{Static analysis}\label{sec:results:static}

\begin{figure}
\centering
\subfloat[]{\includegraphics[height=0.38\textwidth]{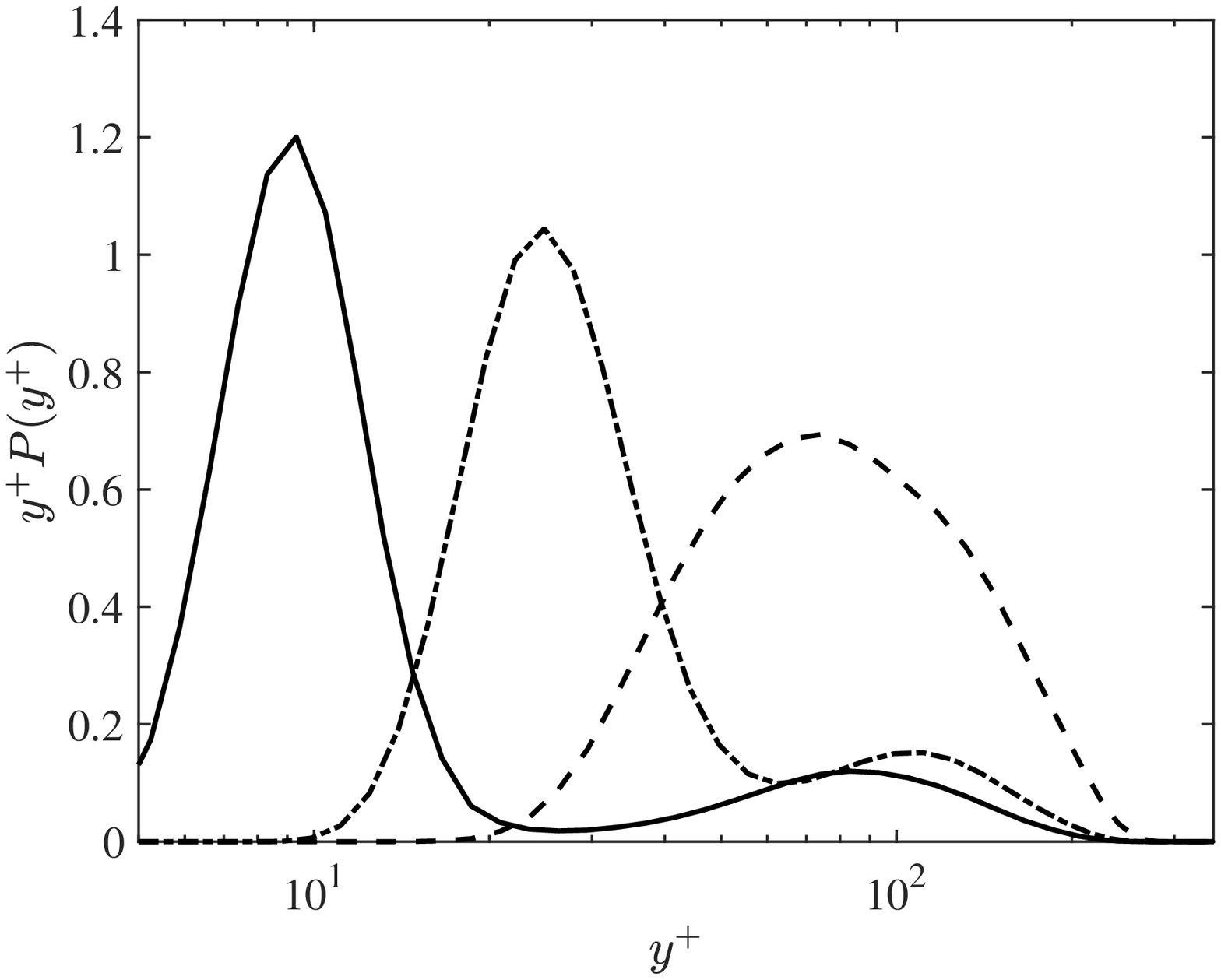}}
\hspace{0.2cm}
\subfloat[]{\includegraphics[height=0.38\textwidth]{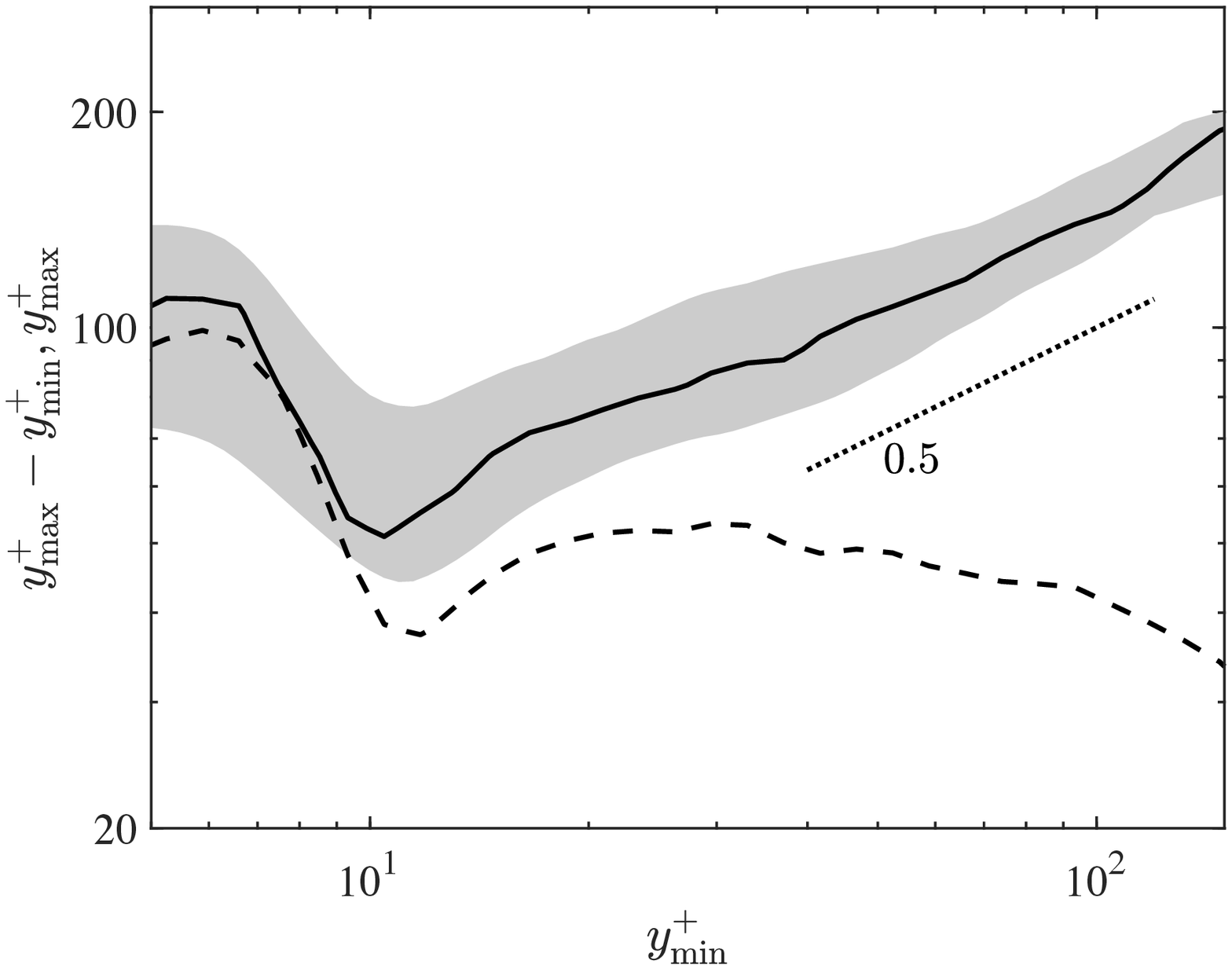}}
\caption{({a}) Premultiplied p.d.f. of $y_{\min}^+$ (solid line), $y_{\max}^+$ (dashed line) and centroid, $y_c$, (dot-dashed line) of the streaks. ({b}) C.d.f. of $y_{\max}^+$ as a function of $y_{\min}^+$; solid line indicates the median quantity and shaded region indicate 25 and 75 percentiles. Dashed line indicates median value of $y_{\max}^+ - y_{\min}^+$ as a function of $y_{\min}^+$. }
\label{fig:y_pdf}
\end{figure}

First, we consider all the nodes of the supergraph as individual entities and perform a static analysis. The goal is to study the overall statistics of streaks. Figure~\ref{fig:y_pdf}({a}) shows the probability density function (p.d.f.), denoted $P(\cdot)$, of the wall-normal locations of the bottom ($y_{\min}$), the top ($y_{\max}$), and the centroid ($y_c$) of the streak structures. As expected, most of the streaks occur close to the wall ($y_{\min}^+ < 20$) with a much smaller secondary peak in the p.d.f of $y_{\min}$ centered around $y^+\simeq 90$. The p.d.f.s of $y_c$ and $y_{\max}$ are centered around $y^+\simeq 20$ and $y^+\simeq 70$, respectively and have a wider spread than that of $y_{\min}$.

The cumulative distribution function (c.d.f.) $y_{\max}$ as a function of $y_{\min}$ in Figure~\ref{fig:y_pdf}({b}) shows a clear divide in streaks that are wall-attached ($y_{\min}^+ < 20$) versus wall-detached ($y_{\min}^+>20$). Wall-detached streaks are positively correlated with the corresponding $y_{\max}^+$ with $y_{\max}^+\sim (y_{\min}^+)^{0.5}$. Thus, the height of the streaks grows only sublinearly with distance to the wall, with the streak height becoming smaller beyond a certain threshold, as shown by the median quantites of $(y_{\max}^+ - y_{\min}^+)$ in figure \ref{fig:y_pdf}(b). However, with wall-attached structures, more structures end with $y_{\max}^+>70$, forming `tall' attached streaks. While larger Reynolds number simulations will have a fully developed logarithmic layer \cite{Lee2015}, the current Reynolds number is too small to identify a distinctive logarithmic layer. Thus, it is difficult to provide a clear cut definition of `tall' structures whose $y_{\max}$ outside the buffer layer; however, we choose the cutoff for `tall' structures to be $y^+_\mathrm{cutoff}=70$ as the two peaks of the bimodal distribution of $y_c$ in Figure~\ref{fig:y_pdf}({a}) are separated at $y^+\approx 70$, providing a natural divide. The $y^+=70$ cutoff also coincides with the location where the dominant production mechanism chanages \citep{Lee2019}. We classify the streaks as wall-detached if $y_{\min}^+>20$, tall attached if $y_{\min}^+<20$ and $y_{\max}^+>70$, and buffer layer if $y_{\min}^+<20$ and $y_{\max}^+<70$ (see Figure~\ref{fig:diagram}a--c). The definitions of the wall-attached/detached streaks follow similar analysis for Reynolds stresses and clusters \cite{delAlamo2006,Lozano-Duran2012,Lozano-Duran2014}. Based on these studies, the attached streaks should behave differently compared to the detached ones \cite{Townsend1961,Townsend1976} with the detached streaks being more isotropic and the attached structures in the log-layer forming self-similar families with approximately constant geometric aspect ratios \cite{Jimenez2012,Jimenez2013}. The current domain lacks a well-defined log layer, and will not contain a distinct layer of self-similar streaks; however, we expect self-similar streaks to be dominant in higher Reynolds number flows. While this threshold for `tall' structures is arbitrary, changing the limit to $y^+_\mathrm{cutoff}=60$ or $80$ does not change the conclusions of this paper, and thus, the results are resilient to the choice of this threshold. 

\begin{figure}
\begin{center}
\subfloat[]{\includegraphics[height=0.32\textwidth]{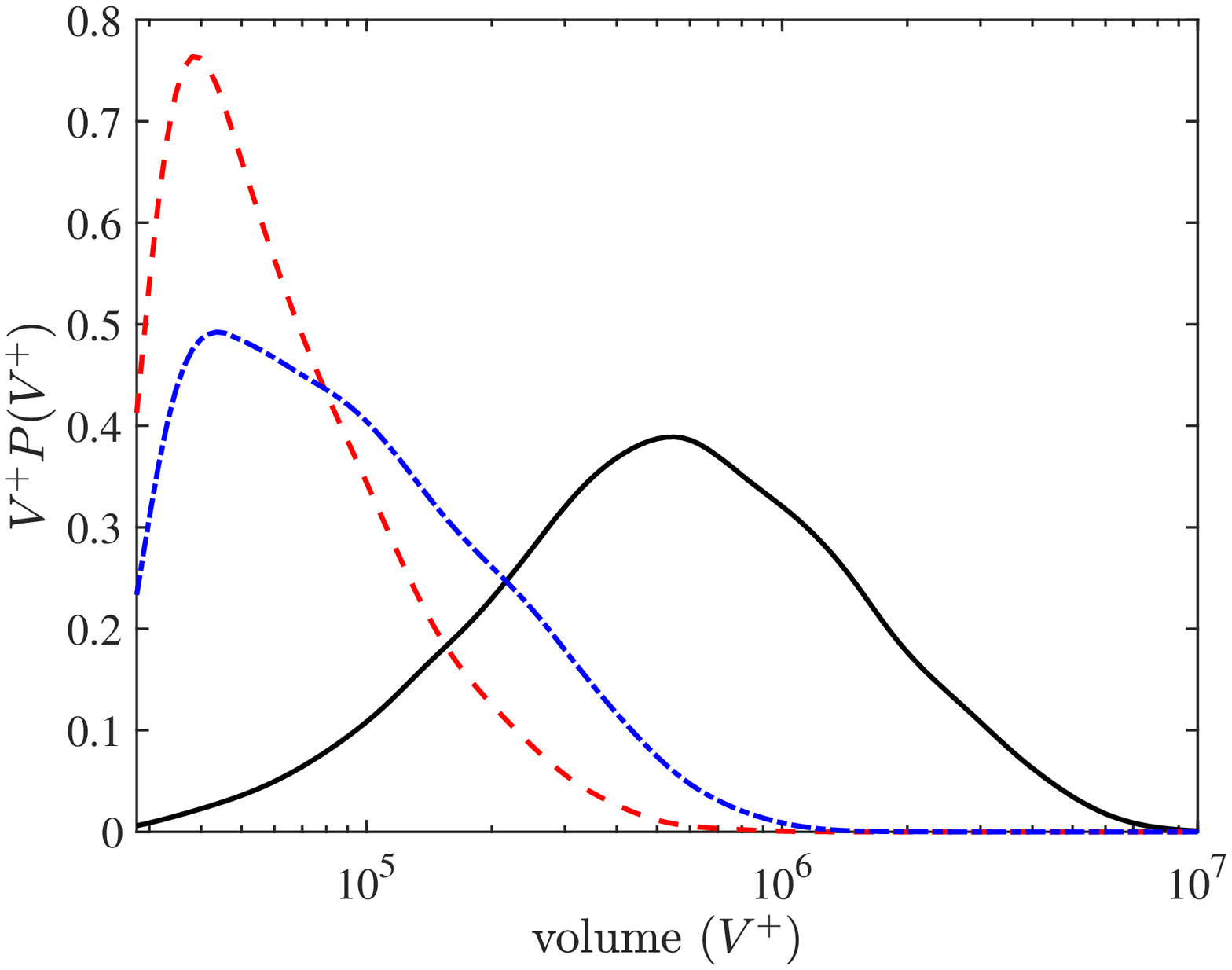}}
\hspace{0.2cm}
\subfloat[]{\includegraphics[height=0.32\textwidth]{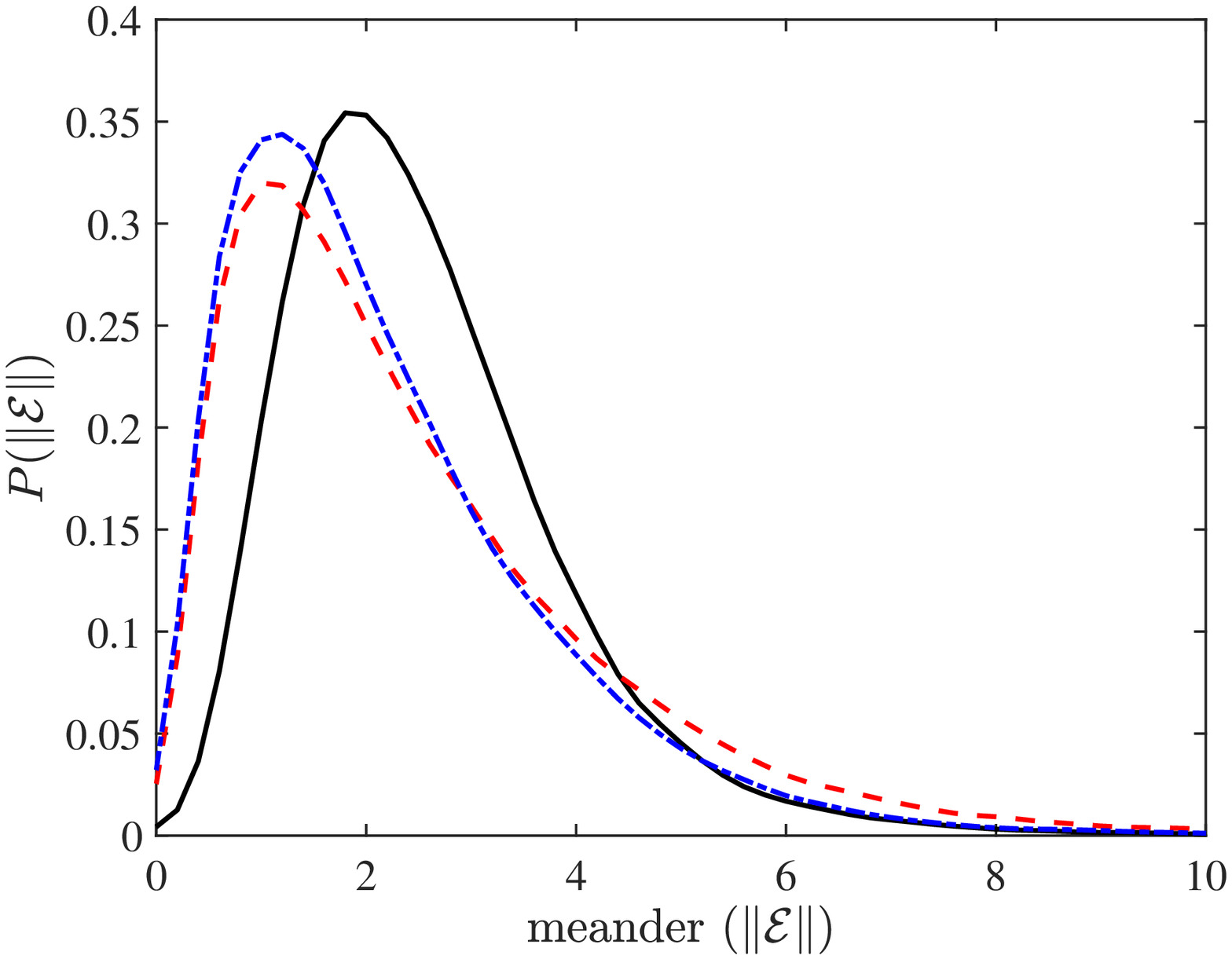}}\\
\subfloat[]{\includegraphics[height=0.32\textwidth]{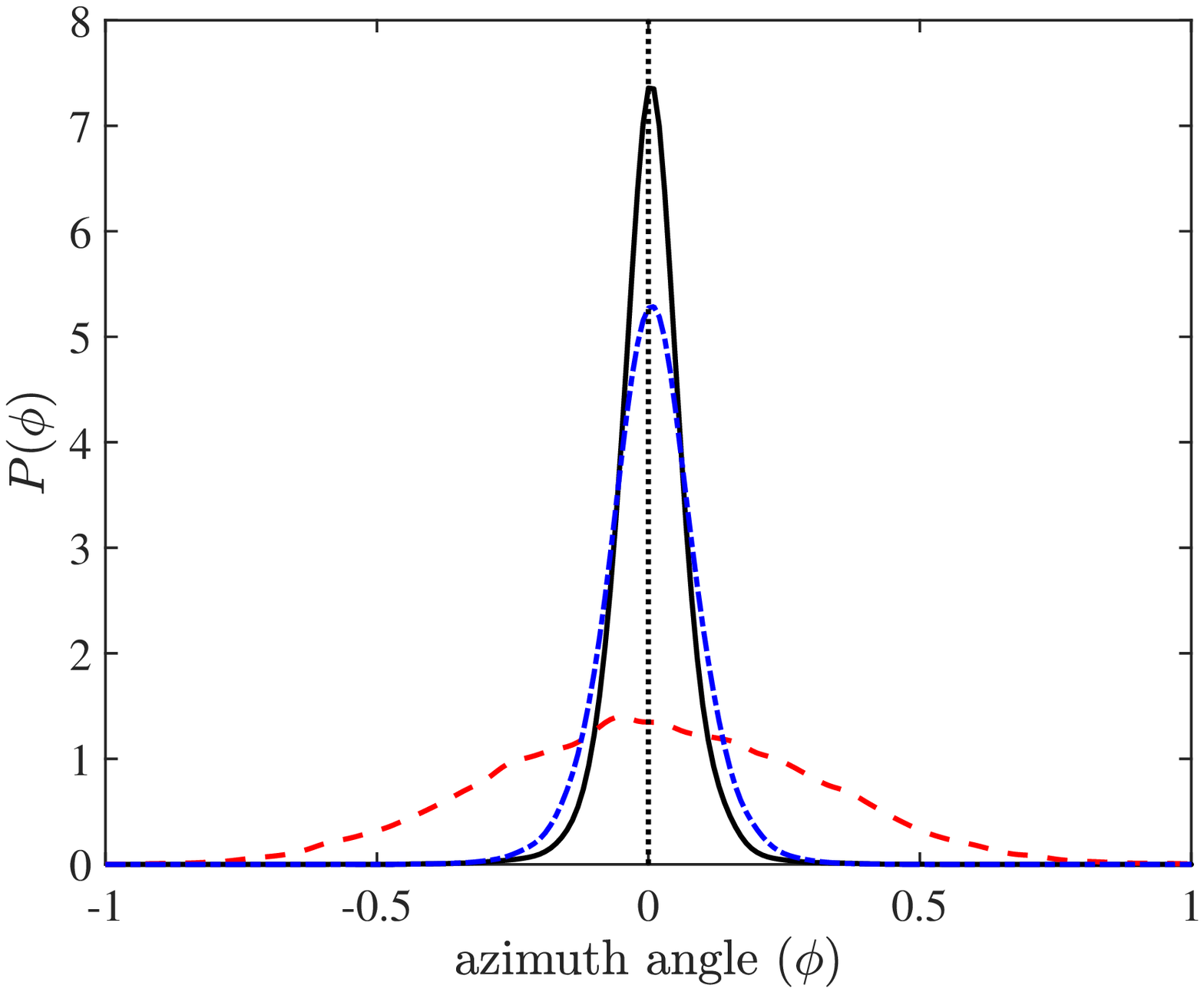}}
\hspace{0.2cm}
\subfloat[]{\includegraphics[height=0.32\textwidth]{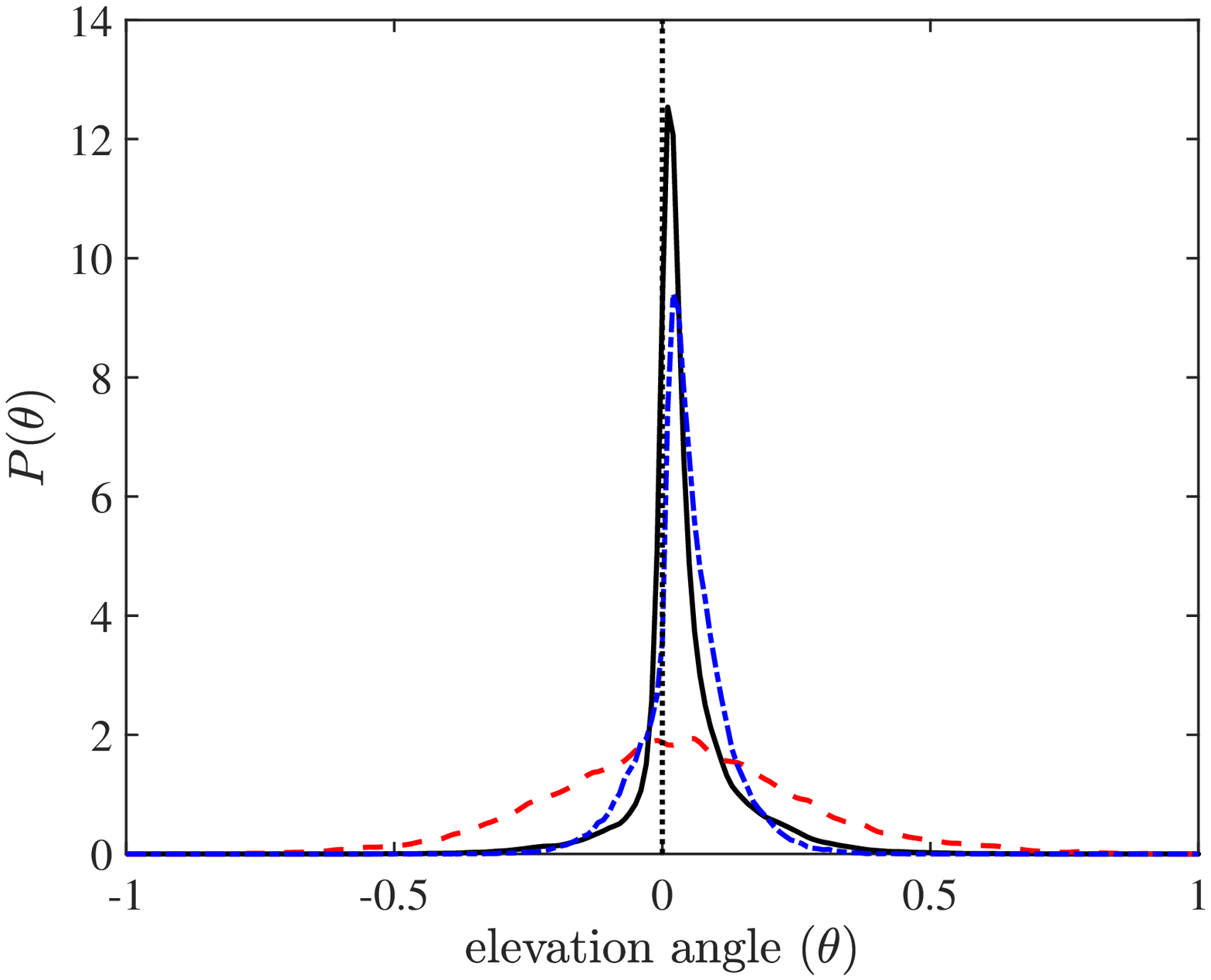}}\\
\subfloat[]{\includegraphics[height=0.32\textwidth]{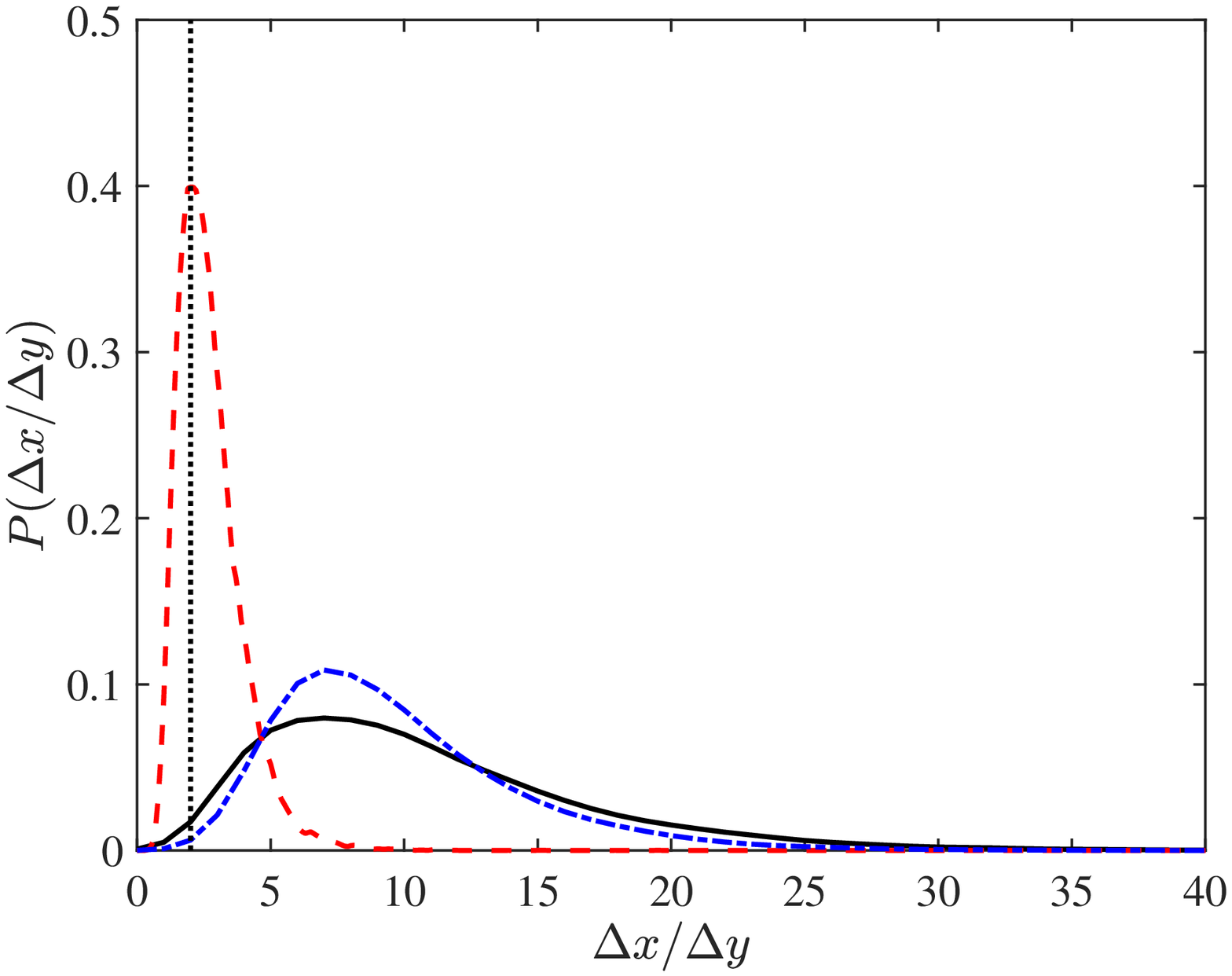}}
\hspace{0.2cm}
\subfloat[]{\includegraphics[height=0.32\textwidth]{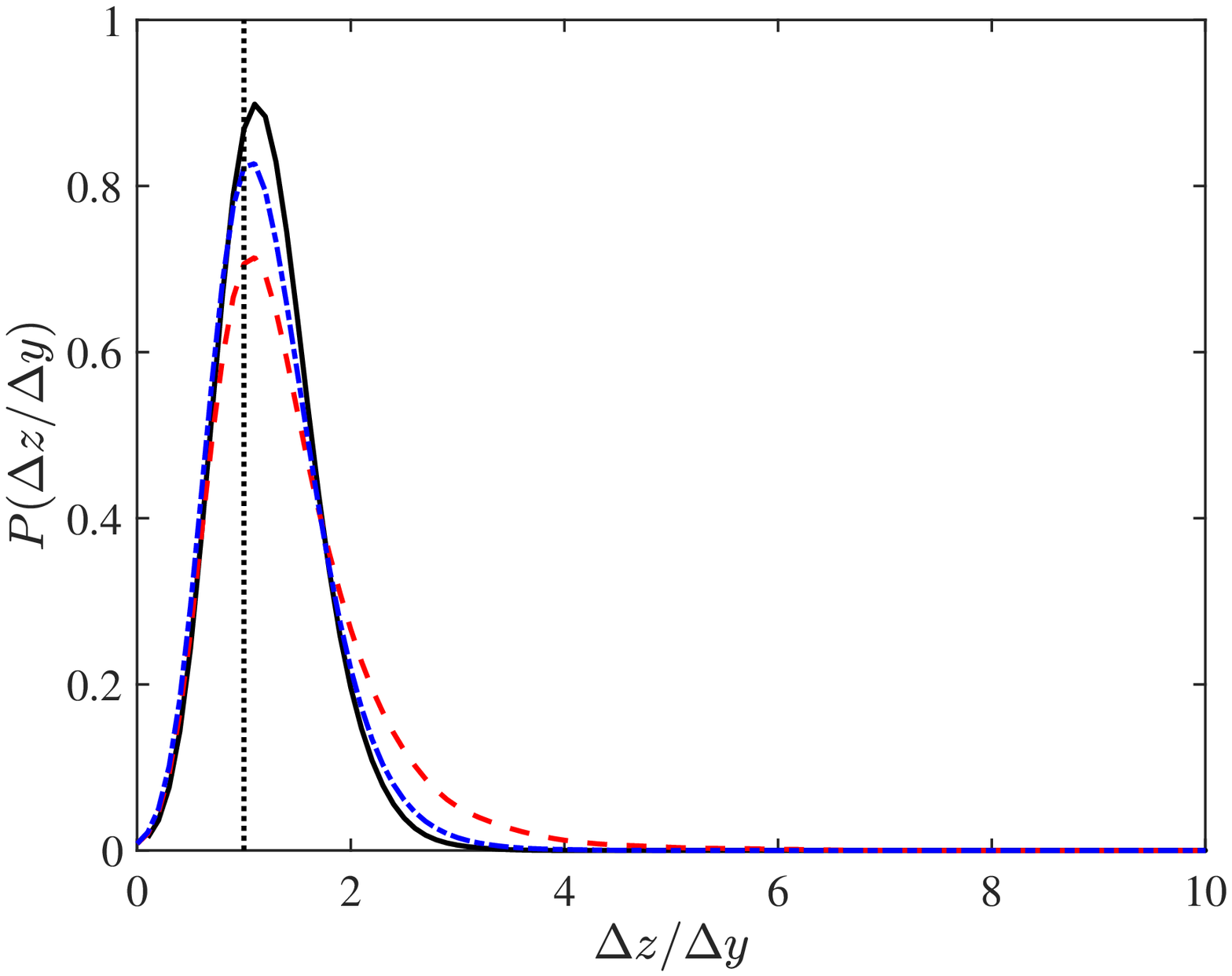}}
\end{center}
\caption{(a) Premultiplied p.d.f. of the  volume and the p.d.f. of ({b}) meandering quantity, ({c}) azimuth and ({d}) elevation angle of the spine, and ({e}) streamwise/wall-normal and ({f}) spanwise/wall-normal aspect ratio of the bounding box. Lines indicate tall attached streaks (black solid line), detached streaks (red dashed line) and buffer layer streaks (blue dot-dashed line). Dotted lines are (c) $\phi = 0$, (d) $\theta= 0$, (e) $\Delta x/\Delta y =2$ and (f) $\Delta x/\Delta z = 1$.}
\label{fig:static_pdf}
\end{figure}
We study the distribution of key statistics of the streaks for the tall-attached, detached, and buffer layer streaks. Of all the identified streaks, the majority are categorized as buffer layer streaks (73.6\%). The tall-attached and detached streaks account for 12.3\% and 14\% of the streaks, respectively. In Figure~\ref{fig:static_pdf}, the p.d.f. of volume, meandering coefficient, azimuth, and elevation angle of the spine, and aspect ratios of the bounding box are shown for the three classes of streaks. The three categories have distinct properties. For example, the tall-attached streaks tend to be larger in volume, meander more, are aligned along the streamwise direction with $\phi\simeq 0$, and are more elongated. This shows that even if the tall-attached streaks meander more, their larger dimensions allow them to align their mean orientation in the streamwise direction, as depicted by their azimuth angle. On the other hand, detached streaks are smaller and more isotropic, and have varying orientations, as shown by the wide range of azimuth and elevation angles. Even with varying orientations, the size of these detached streaks is too small for them to meander more. Finally, buffer layer streaks are smaller and aligned along the streamwise direction, meander less than tall-attached streaks, and are elongated only in the streamwise direction. These are the organized narrow streamwise-elongated streaks observed in the buffer layer \cite{Kline1967}. Still, all three types of streaks have a similar wall-normal/spanwise aspect ratio of approximately unity and skewness of $\phi\approx 0$ and $\theta > 0$, indicating, on average, a positive tilt.

\begin{figure}
\begin{center}
\subfloat[]{\includegraphics[height=0.37\textwidth]{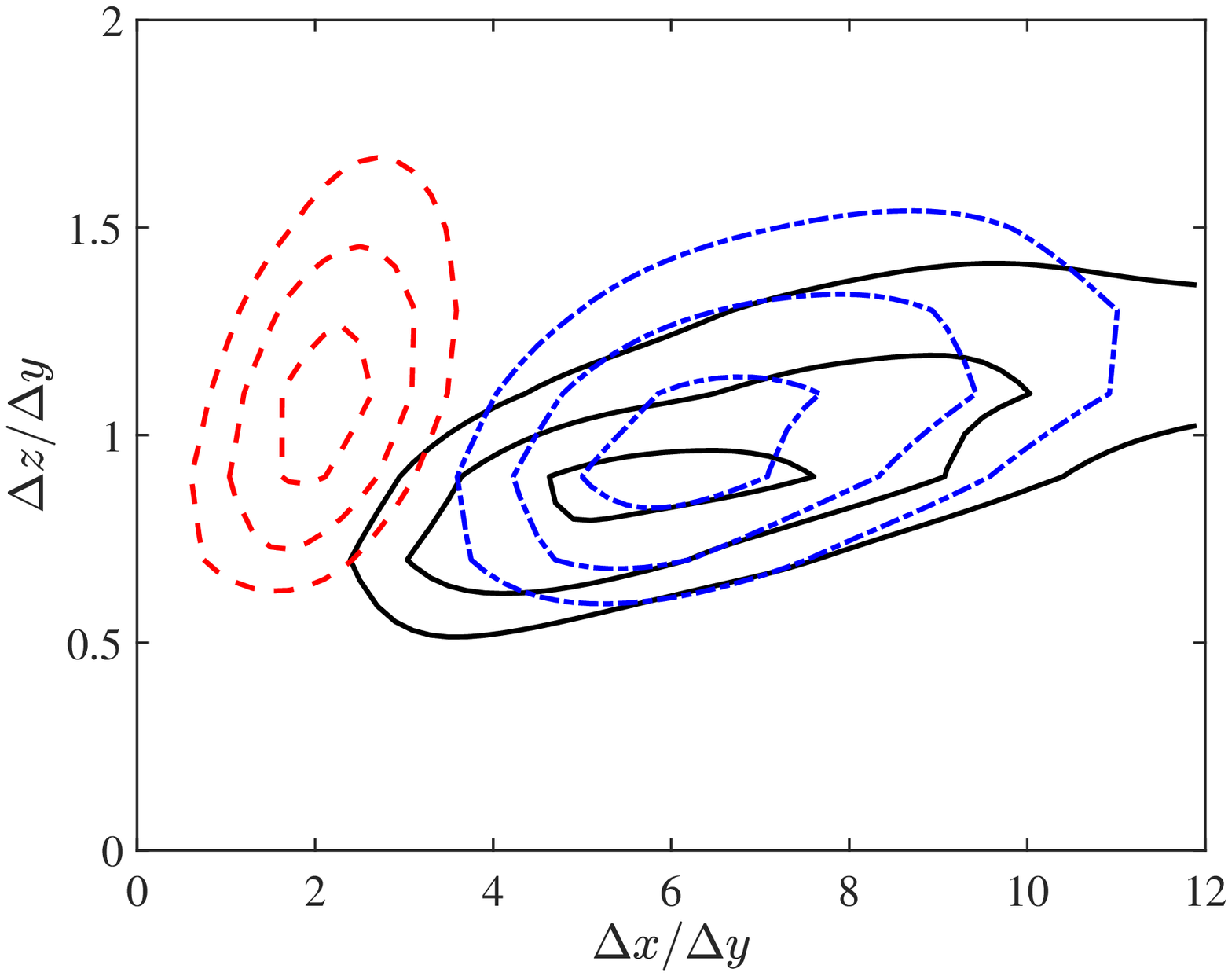}}
\hspace{0.2cm}
\subfloat[]{\includegraphics[height=0.37\textwidth]{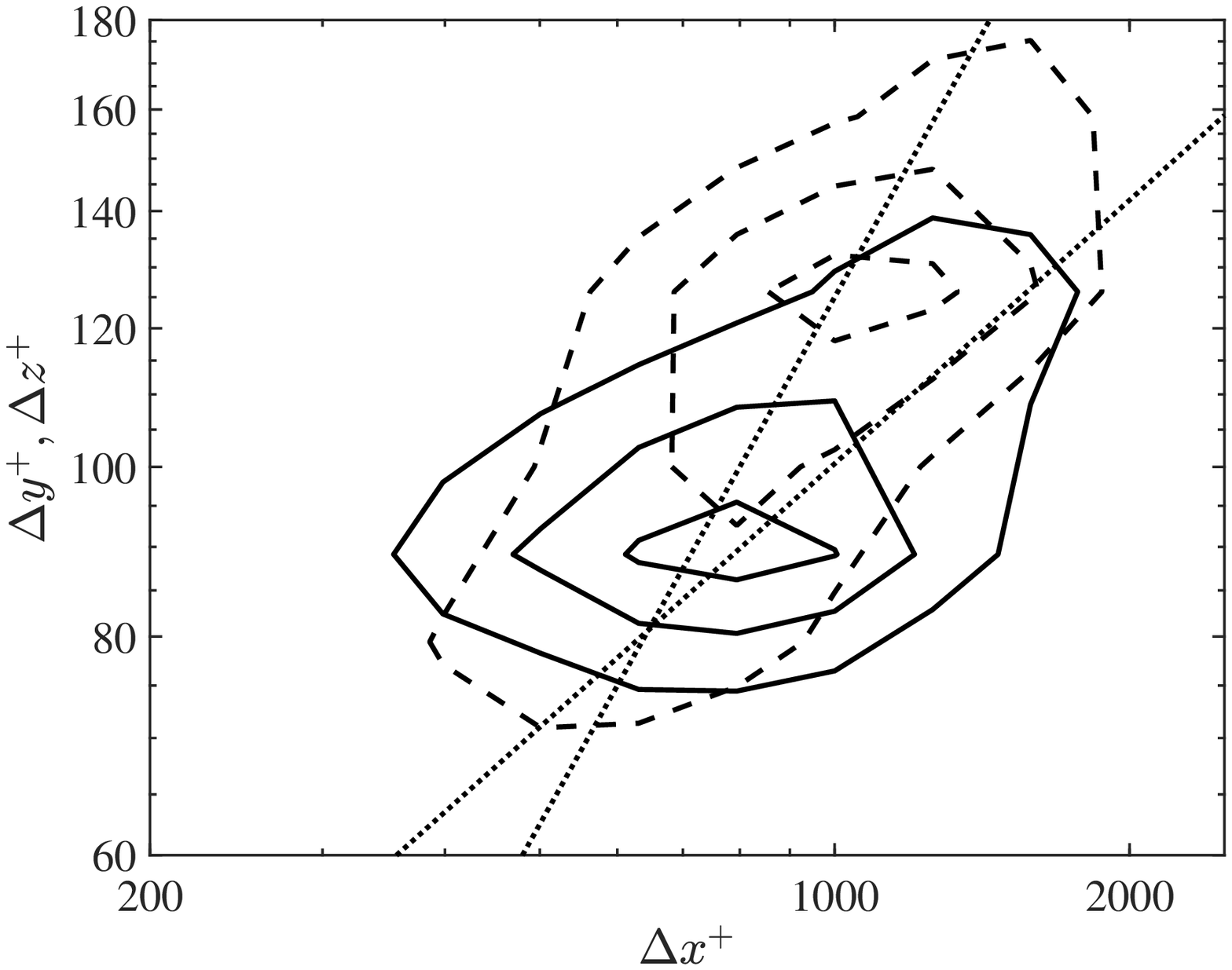}}\\
\subfloat[]{\includegraphics[height=0.37\textwidth]{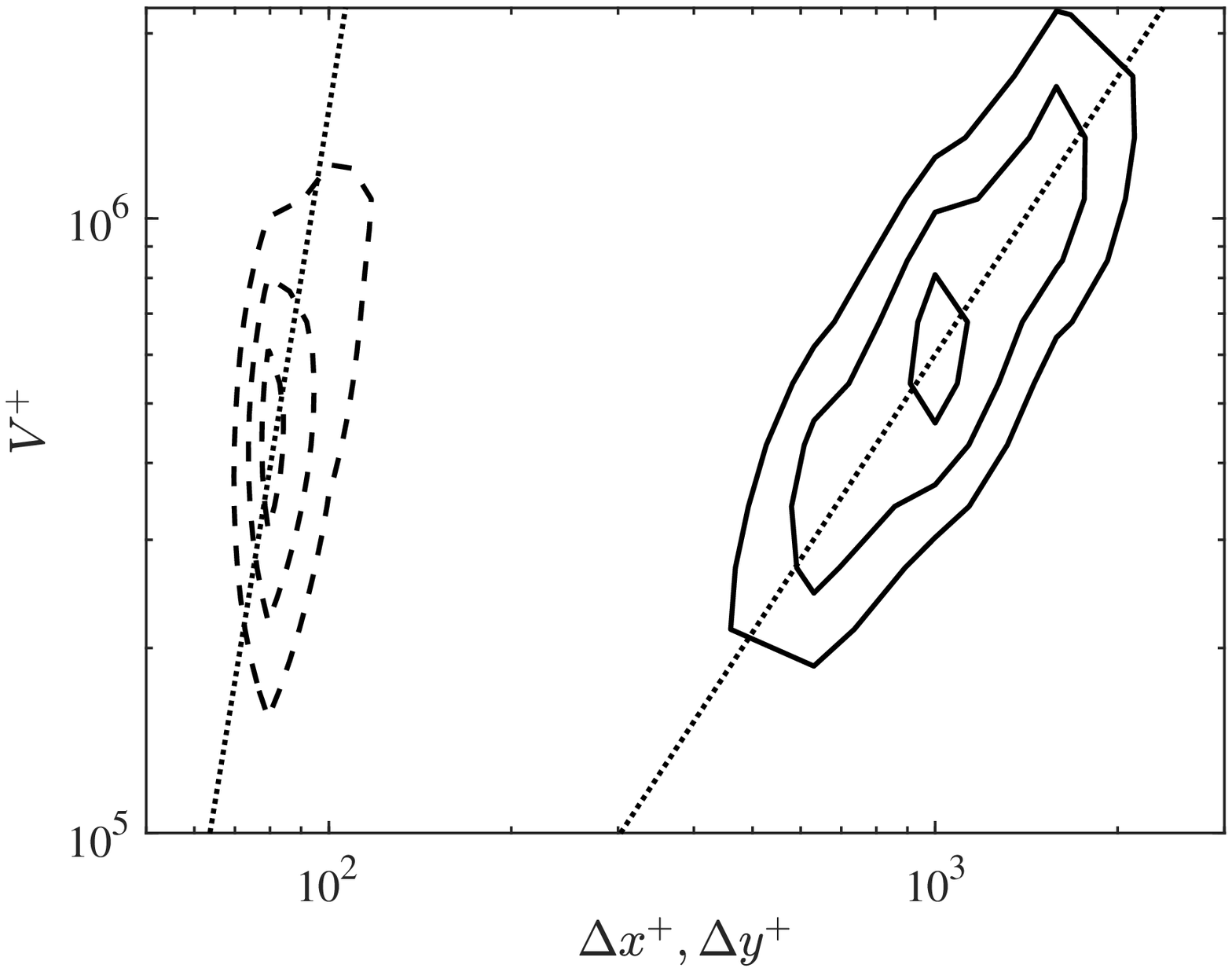}}
\hspace{0.2cm}
\subfloat[]{\includegraphics[height=0.37\textwidth]{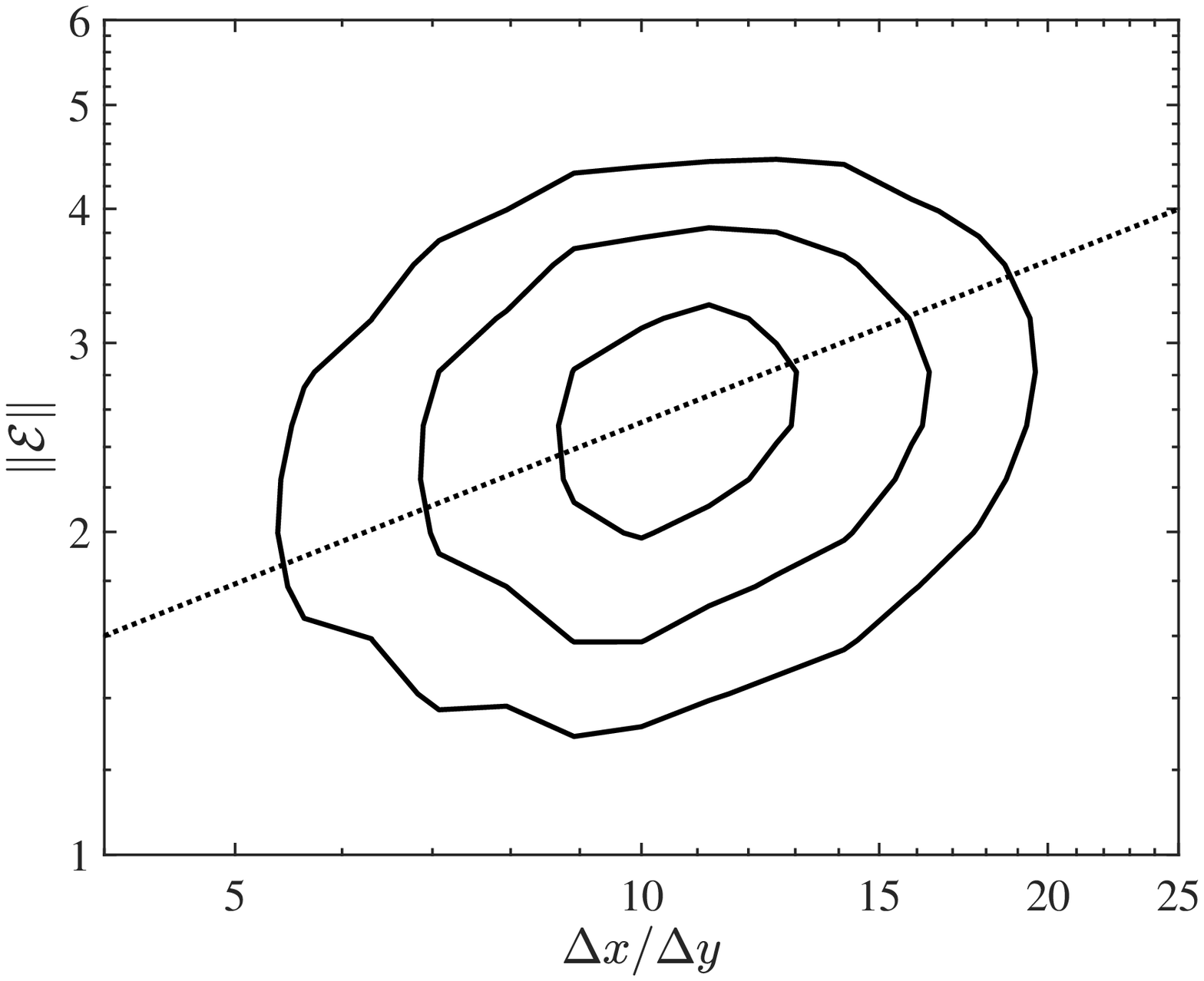}}
\end{center}
\caption{({a}) Joint p.d.f. of the aspect ratios $\Delta x/\Delta y$ and $\Delta z/\Delta y$ for tall attached streaks (black solid line), detached streaks (red dashed line) and buffer layer streaks (blue dot-dashed line). ({b}) Joint p.d.f. of the logarithms of dimensions of the bounding box $\Delta x$ with $\Delta y$ (solid lines) and $\Delta z$ (dashed lines) for tall-attached streaks. Dotted line indicates $\Delta x\sim \Delta y^2$ and $\Delta x\sim \Delta z$. ({c}) Joint p.d.f. of the logarithms of volume with $\Delta x$ (solid lines) and $\Delta y$ (dashed lines) for tall-attached streaks. Dotted lines are $V \sim\Delta x^{3/2}$ and $V\sim\Delta y^6$. ({d}) Joint p.d.f. of the logarithms of meandering quantity and $\Delta x/\Delta y$ for tall attached streaks. Dotted line is $\|\mathcal{E}\|\sim (\Delta x/\Delta y)^{1/2}$. Contour levels are 50, 70, 90\% of maximum value.}
\label{fig:static_jpdf}
\end{figure}

The size of the average bounding box is $5.12\delta\times0.49\delta\times0.58\delta$ for the tall attached streaks, $0.62\delta\times0.26\delta\times0.33\delta$ for the detached streaks, and $1.93\delta\times0.21\delta\times0.27\delta$ (or $300\nu/u_\tau \times 38\nu/u_\tau \times 48\nu/u_\tau$) for the buffer layer. The tall-attached streaks and the buffer layer streaks have similar aspect ratios (Figure~\ref{fig:static_jpdf}{a}), although the trend deviates at high values of $\Delta x/\Delta y$. These aspect ratios of the attached streaks are similar to what other people observe for buffer layer streaks \cite{Sillero2014}. The streamwise and wall-normal dimensions of the bounding box for tall-attached flows follows $\Delta x \sim \Delta y^2$ and $\Delta x \sim \Delta z$ (Figure~\ref{fig:static_jpdf}{b}), which indicates that the streaks grow in the streamwise and spanwise direction at a faster rate than the wall-normal direction. 
Detached streaks are more isotropic in size. They are also less aligned with the streamwise direction and might be the reason that in the outer region, oblique features become more prominent \cite{Kevin2019}. The volume of the tall-attached streaks follow $V\sim \Delta x^{3/2}, \Delta y^6$ (Figure~\ref{fig:static_jpdf}{c}), showing that the volume of the bounding box ($\Delta x \Delta y \Delta z \sim \Delta x^{5/2}, y^5$) is not a good representation of the volume of the streak, most likely due to a large amount of meandering, especially for large streaks. We also see that meandering intensifies with increasing streak volume, albeit weakly, following $\|\mathcal{E}\|\sim (\Delta x/\Delta y)^{1/2}$. This indicates that meandering is a byproduct of the elongating (and enlarging) streak, which makes the streak less likely to orient in the same direction throughout the entirety of its structure.

%

\section{Temporal analysis}\label{sec:results:temp}

\begin{figure}
\begin{center}
\subfloat[]{\includegraphics[height=0.38\textwidth]{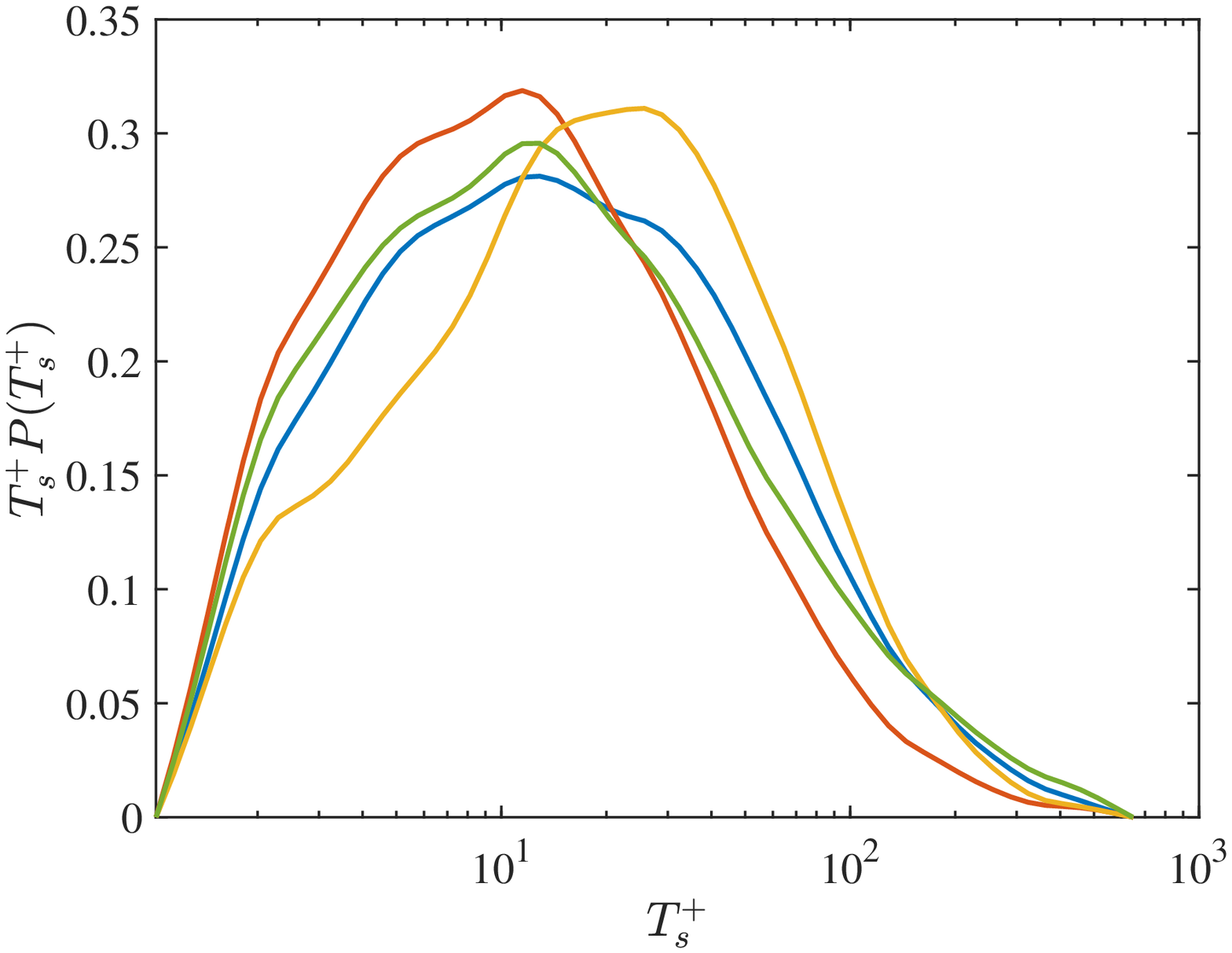}}
\hspace{0.2cm}
\subfloat[]{\includegraphics[height=0.38\textwidth]{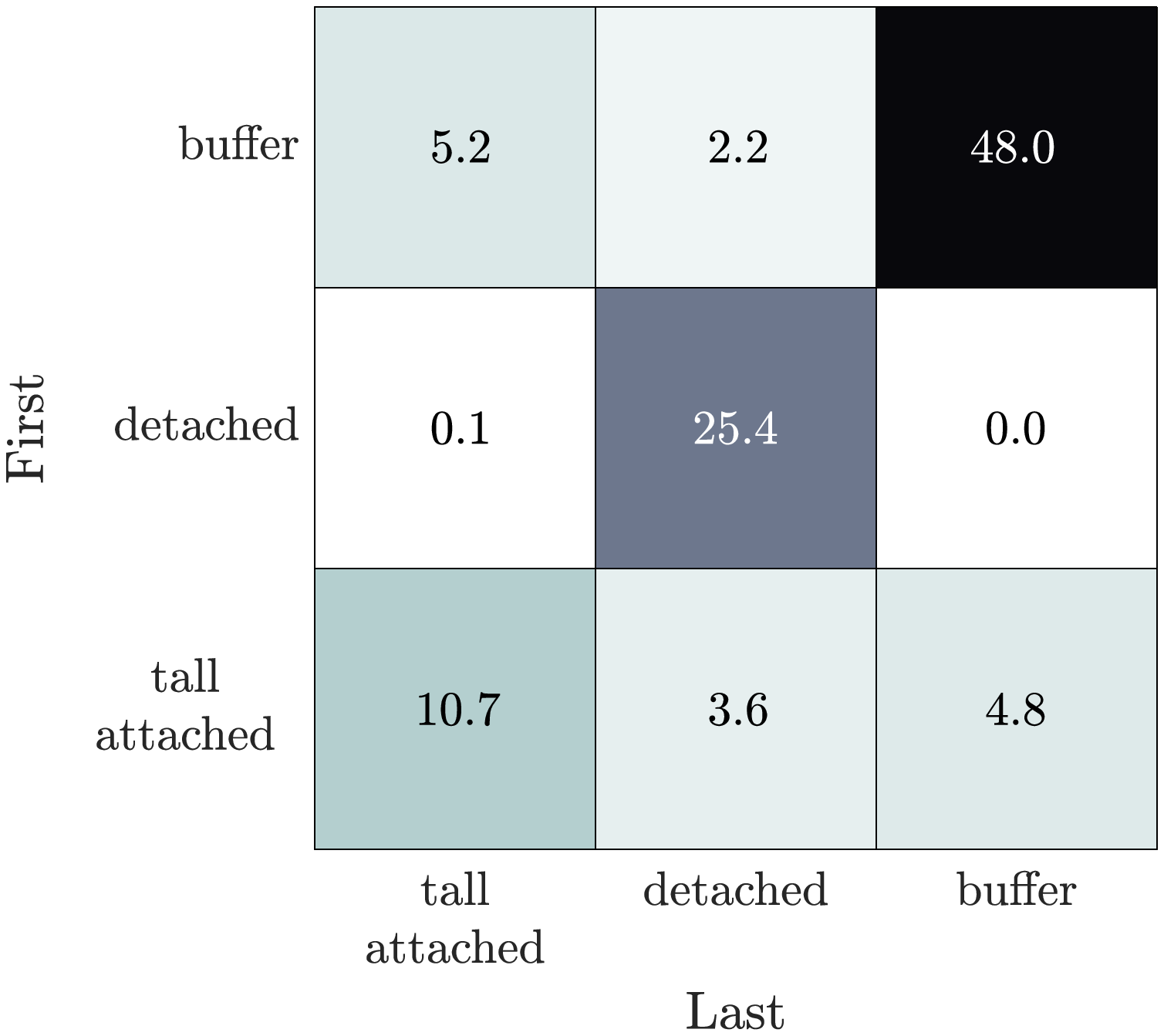}}
\end{center}
\caption{(a) Premultiplied p.d.f. of lifetimes of streaks for primary (yellow), incoming (blue), outgoing (red), and connector (green) branches. (b) Percentage of branches starting from (first) and ending in (last) each streak category.}
\label{fig:evolution}
\end{figure}

We now consider each branch (individual structures within each graph) as one entity, providing temporal analysis of the life-cycle of streaks. The p.d.f. of the lifetimes, $T_s$, for the different branch types are given in Figure~\ref{fig:evolution}(a). All branches have a long tail, with the longest lifetime being $T_s^+ \approx 600$. However, the median lifetime is less than 20$\nu/u_\tau^2$, with the mean lifetime being longest for the main branches (32.1$\nu/u_\tau^2$) and shortest for outgoing branches (21.5$\nu/u_\tau^2$). This shows that the current temporal domain is long enough to analyze the full life-cycle of streaks, including the longest ones observed. 

We analyze how the streaks change within each branch. The evolution of a branch defined by the classification of its first and last streak (node) in time is summarized in Figure~\ref{fig:evolution}(b). Majority of the branches stay in the same category they were created in -- either as tall-attached, detached, or buffer layer streaks. Only a small percentage (8\%) of the branches created as wall-attached streaks, either tall or buffer layer, detach from the wall with time. On the contrary, a large portion (22\%) of the branches ending as wall-detached streaks were initially wall-attached. Less than 0.1\% of the streaks are created as wall-detached and then attach to the wall. This indicates that the branches that start as wall attached can, albeit rarely, detach and form detached streaks, but not the other way around. The most frequent change in classification is between tall-attached and buffer layer streaks, which occur in similar numbers in either direction; however, this is expected as the $y^+=70$ cutoff for `tall' structures are arbitrary and fluctuations of the streak height about this cutoff will result in a change in classification. 

We also study the classification of the first and last streak (node) in a branch for each branch type (Table~\ref{tab:branch}). Branches created from a turbulent background (primary and incoming branches) rarely contain tall-attached streaks because there is not enough time for the streaks generated this way to grow into tall-attached structures before they merge into another streak or disappear. The majority of the branches created this way start as buffer-layer streaks, accounting for 95\% of these cases. Similarly, branches disintegrating to the turbulent background rarely contain tall-attached streaks, mostly because the streaks break down into smaller streaks before dying. Near the end of the streak life-cycle (last streak of outgoing branches and primary branches), the majority of these streaks are detached streaks, with 57\% being detached and 43\% being buffer layer streaks. Tall-attached streaks appear mostly in the connector branches. 

\begin{table}
\caption{Percent distribution of the first and last streak category for each branch type.}
\centering
\begin{ruledtabular}
\begin{tabular}{ccccc}
 &                 & tall-attached	& detached & buffer layer \\
\colrule
\multirow{4}{*}{First streak} 
 & incoming branch & 0.7            & 2.1      & 97.2         \\
 & outgoing branch & 13.3           & 55.6     & 31.0         \\
 & connector branch& 34.9           & 10.4     & 54.7         \\
 & primary branch  & 1.0            & 9.6      & 89.3         \\
\colrule
\multirow{4}{*}{Last streak}
 & incoming branch & 8.7            & 2.3      & 89.0         \\
 & outgoing branch & 2.9            & 69.5     & 29.6         \\
 & connector branch& 33.2           & 11.8     & 55.0         \\
 & primary branch  & 3.2            & 16.7     & 80.1         \\
\end{tabular}
\label{tab:branch}
\end{ruledtabular}
\end{table}

\begin{figure}
\begin{center}
\subfloat[]{\includegraphics[height=0.38\textwidth]{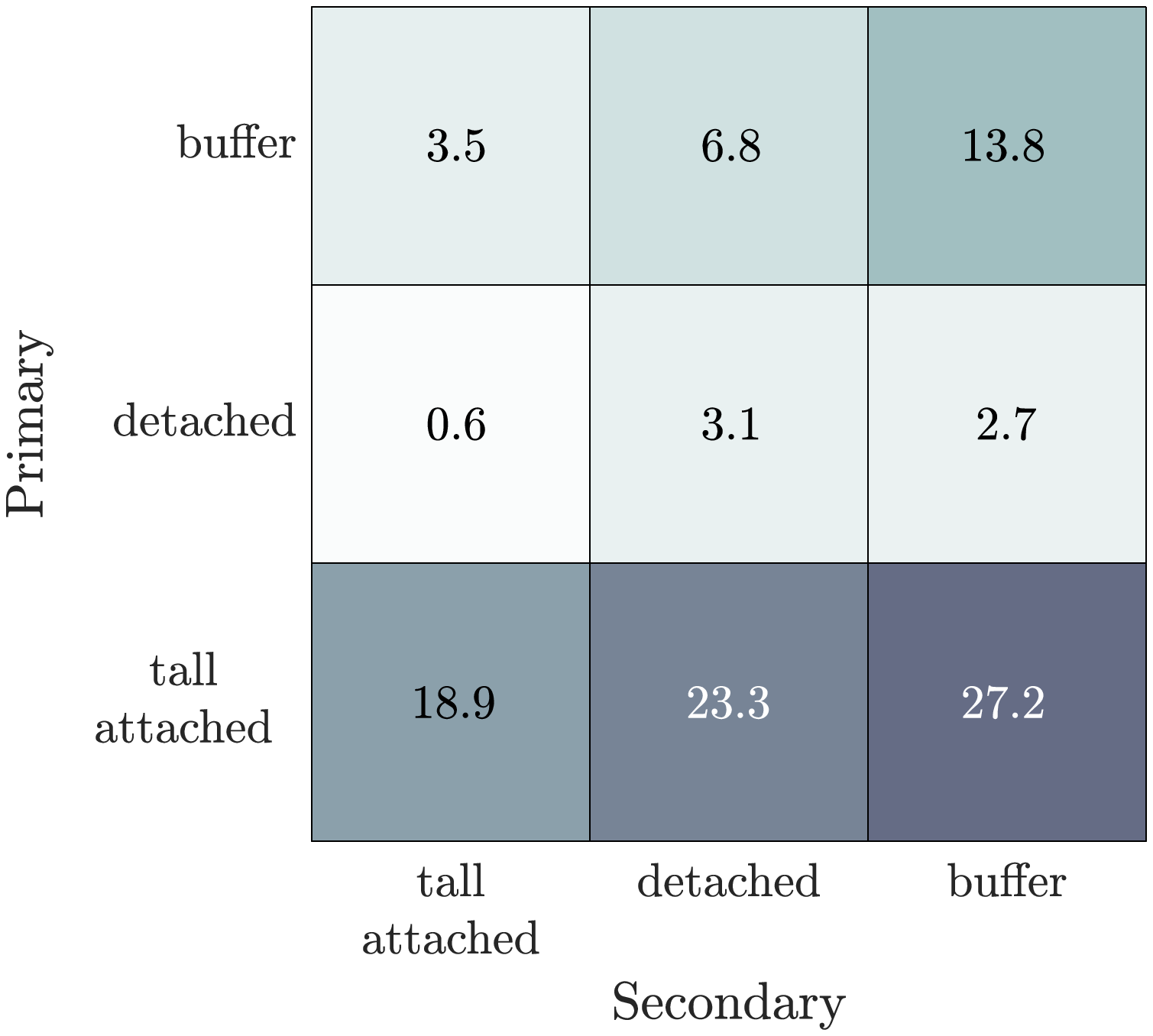}}
\hspace{0.2cm}
\subfloat[]{\includegraphics[height=0.38\textwidth]{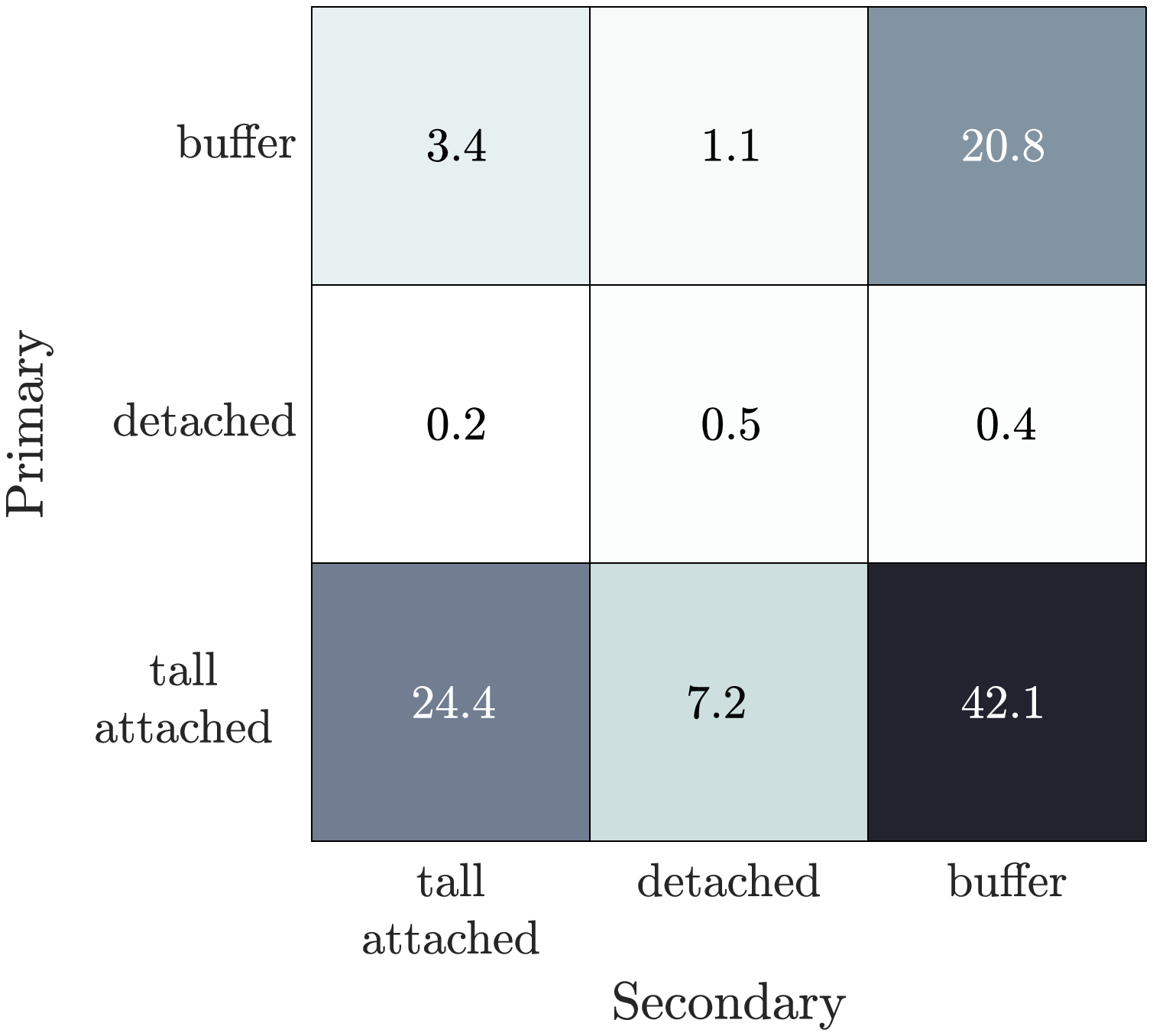}}
\end{center}
\caption{Percentage of primary and secondary streaks in ({a}) split and ({b}) merger events. The secondary streaks split off the primary streaks in a split event, and the secondary streaks merge into the primary streaks in a merger event.}
\label{fig:split_merger}
\end{figure}

The interaction of two branches that either split from or merge into each other can be studied by observing the composition of the primary and secondary structures of each split and merger event, where the primary structure is a node of the main branch in these events. For each split event, the two streaks with the lowest weight, $\Delta V/V_o$, emerging from the same streak are analyzed in Figure~\ref{fig:split_merger}({a}). In most cases, the primary structure is a tall-attached streak with similar probabilities for tall-attached, detached, or buffer layer streak splitting off as the secondary structure. Buffer layer streaks splitting into two buffer layers streaks are the next common. Detached streaks rarely are the primary structures in splitting events. In the case of merger events, the two streaks with the lowest weight merging into the same streak are analyzed in Figure~\ref{fig:split_merger}({b}). The buffer layer streak merging into a tall-attached streak accounts for 42\% of all merger events. The tall-attached streaks merging into tall-attached streaks and buffer layer streaks merging into buffer layer streaks also play a significant role. Detached streaks only occasionally merge into tall-attached streaks. 

The observation provided above shows that the majority of the streaks start as buffer layer streaks and stay as buffer layer streaks. These buffer layer streaks then either die or merge into other existing streaks, which in turn may become tall-attached streaks. Tall-attached streaks are maintained through a continuous coalescing of other buffer-layer or tall-attached streaks. Near the end of their life-time, these streaks disintegrate into detached and buffer-layer streaks, which then dissolves into the turbulent background. This observation is in line with previous works on interaction between large-scale
structures and near-wall structures in a streamwise-minimal channel \cite{Toh2005,Abe2018}. The disintegration into detached and buffer-layer streaks can be thought of as the bursting phenomena, where the production of turbulence in the boundary layer via violent outward eruptions of near-wall fluid. A more detailed analysis of detached streaks and their connection to bursting is given later in this section.

\begin{figure}
\centering
\subfloat[]{\includegraphics[height=0.38\textwidth]{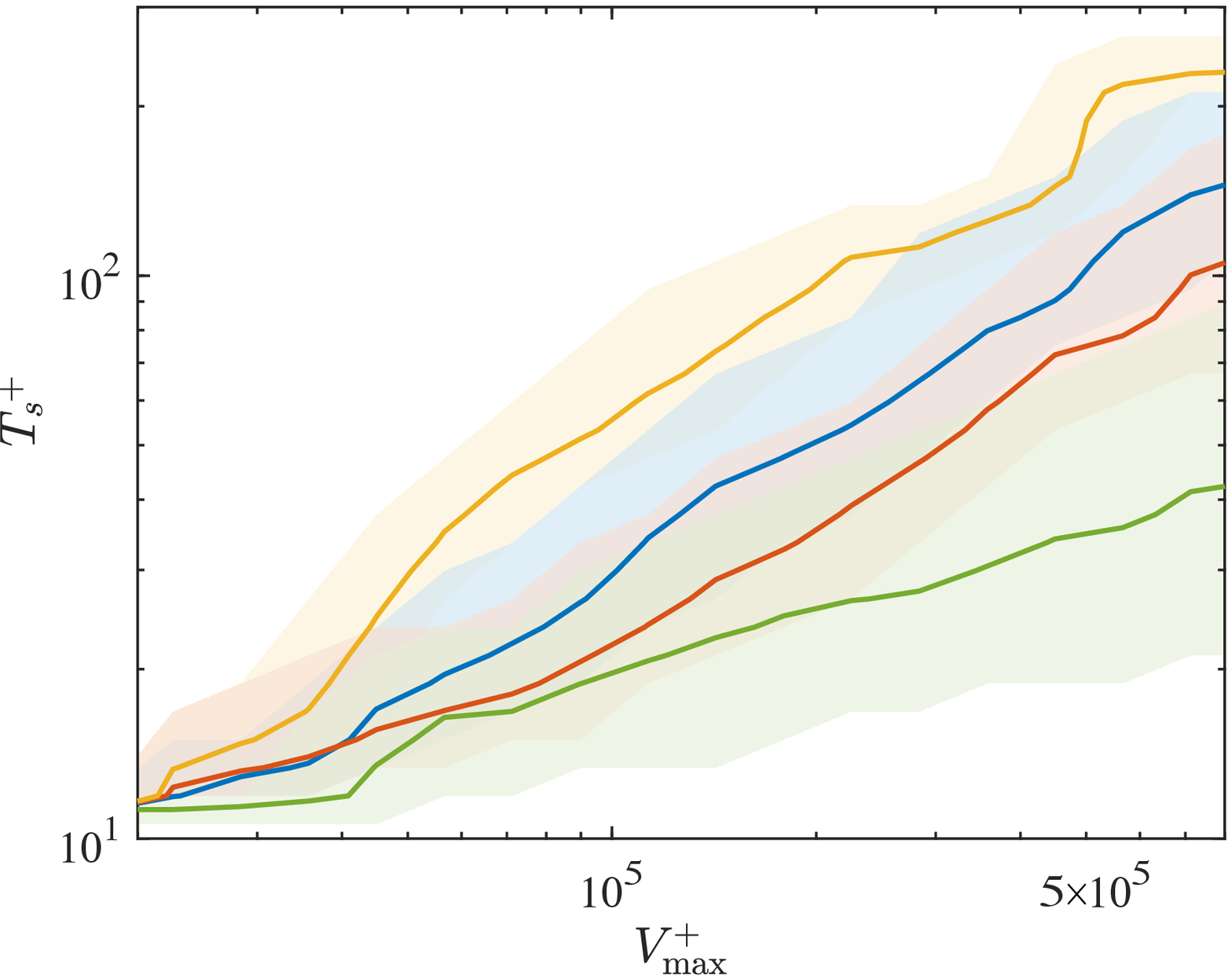}}
\hspace{0.2cm}
\subfloat[]{\includegraphics[height=0.38\textwidth]{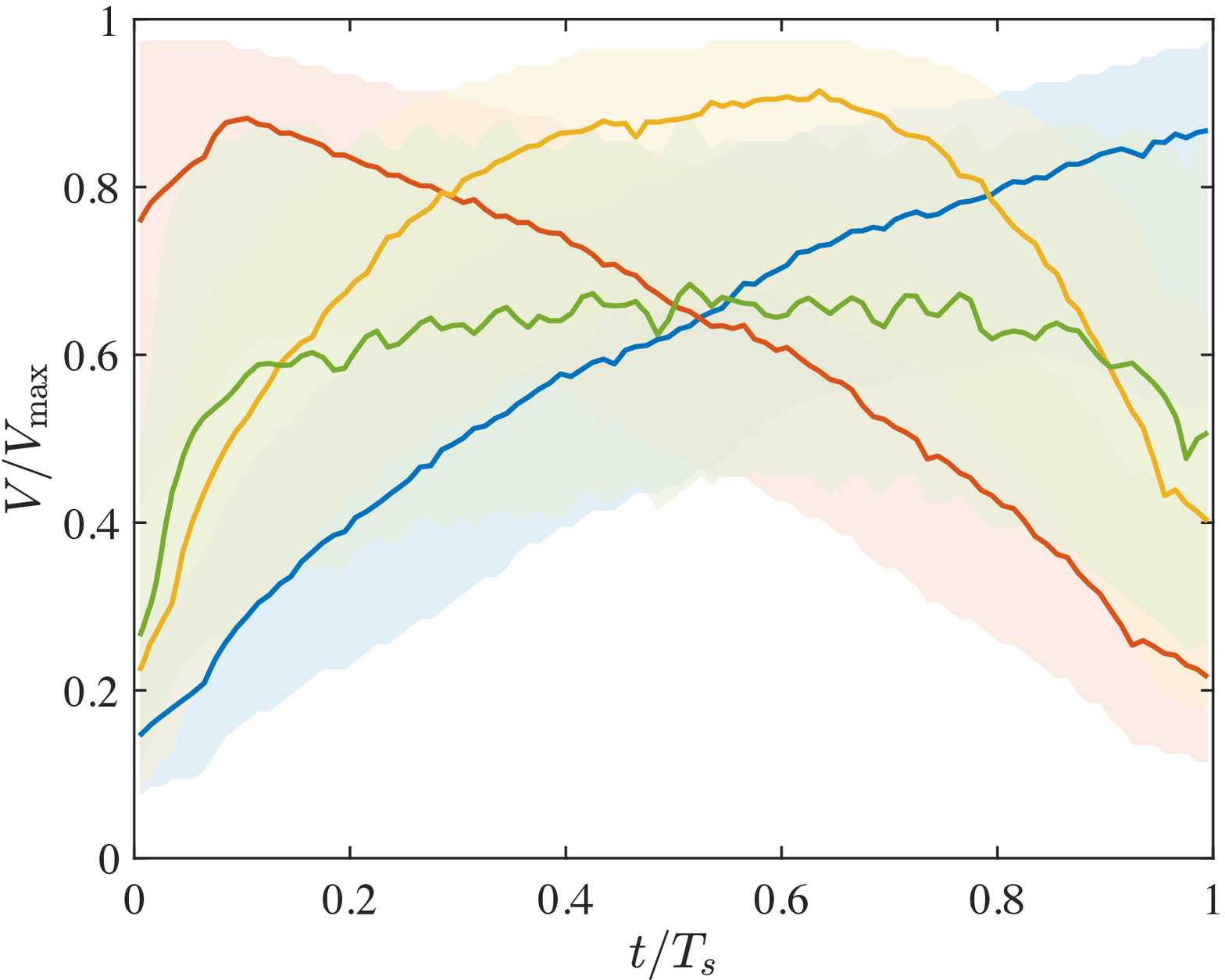}}
\caption{({a}) C.d.f. of the lifetime of a streak as a function of maximum volume of a branch. ({b}) C.d.f. of normalized volume as a function of the lifetime of a streak. Lines indicate median values of primary (yellow), incoming (blue), outgoing (red), and connector (green) branches. Transparent regions are 25 and 75 percentiles.}
\label{fig:volume}
\end{figure}

The lifetimes, $T_s$, of the streaks are correlated with their maximum volume over time, as seen in Figure~\ref{fig:volume}({a}). In all types of branches, a larger maximum volume correlates with larger lifetimes. In the case of incoming branches, the maximum volume occurs near the end of the life-cycle, when it merges into another branch, as seen in Figure~\ref{fig:volume}({b}), indicating the longer the streak is sustained, the larger it is able to grow. For outgoing branches, the maximum volume occurs near the beginning of its life-cycle, where it splits off from another branch, indicating streaks with larger volumes take longer to disintegrate. For primary branches, the volume grows and decays within the streak's lifetime, as expected. Finally, for connector branches, the maximum volume occurs equally likely throughout its lifetime, which explains the smallest correlation between lifetimes and volume for these types of branches in Figure~\ref{fig:volume}({a}). Interestingly, even though meandering correlates with volume (see Figure~\ref{fig:static_jpdf}d for indirect comparison with aspect ratio $\Delta x/\Delta y$), statistically, meandering is not affected by the time evolution of the streaks regardless of branch type (not shown).

\begin{figure}
\centering
\subfloat[]{\includegraphics[height=0.38\textwidth]{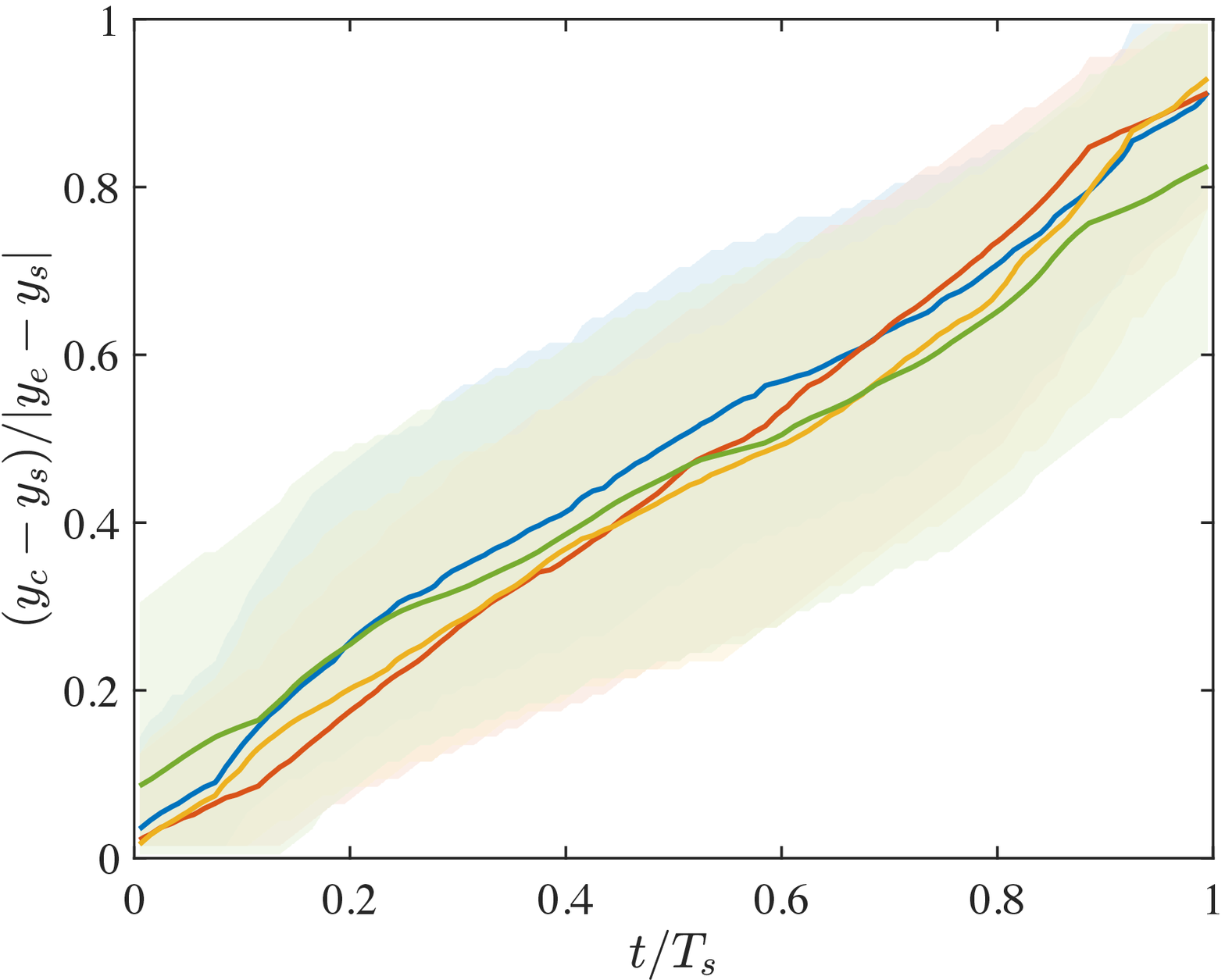}}
\hspace{0.2cm}
\subfloat[]{\includegraphics[height=0.38\textwidth]{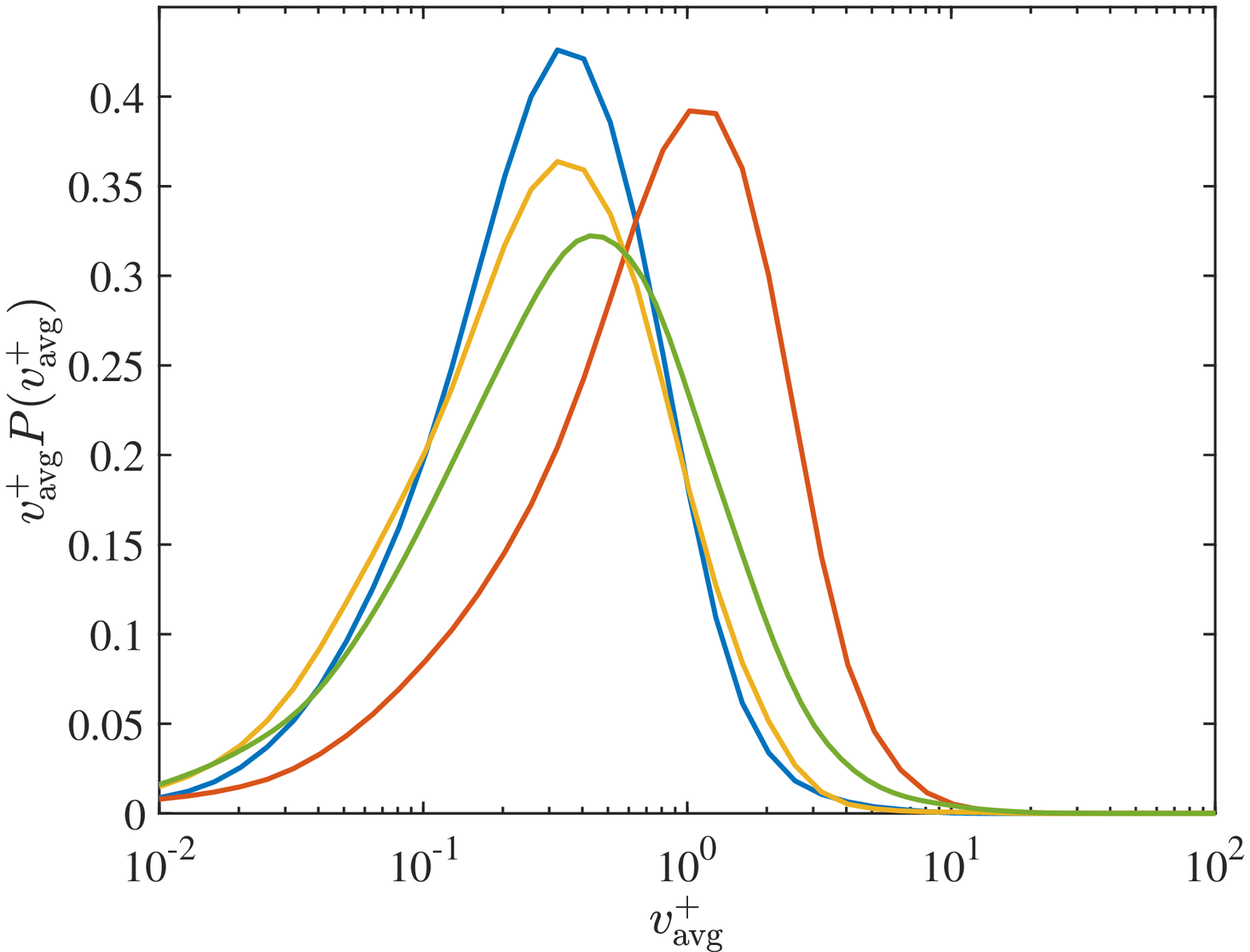}}\\
\subfloat[]{\includegraphics[height=0.38\textwidth]{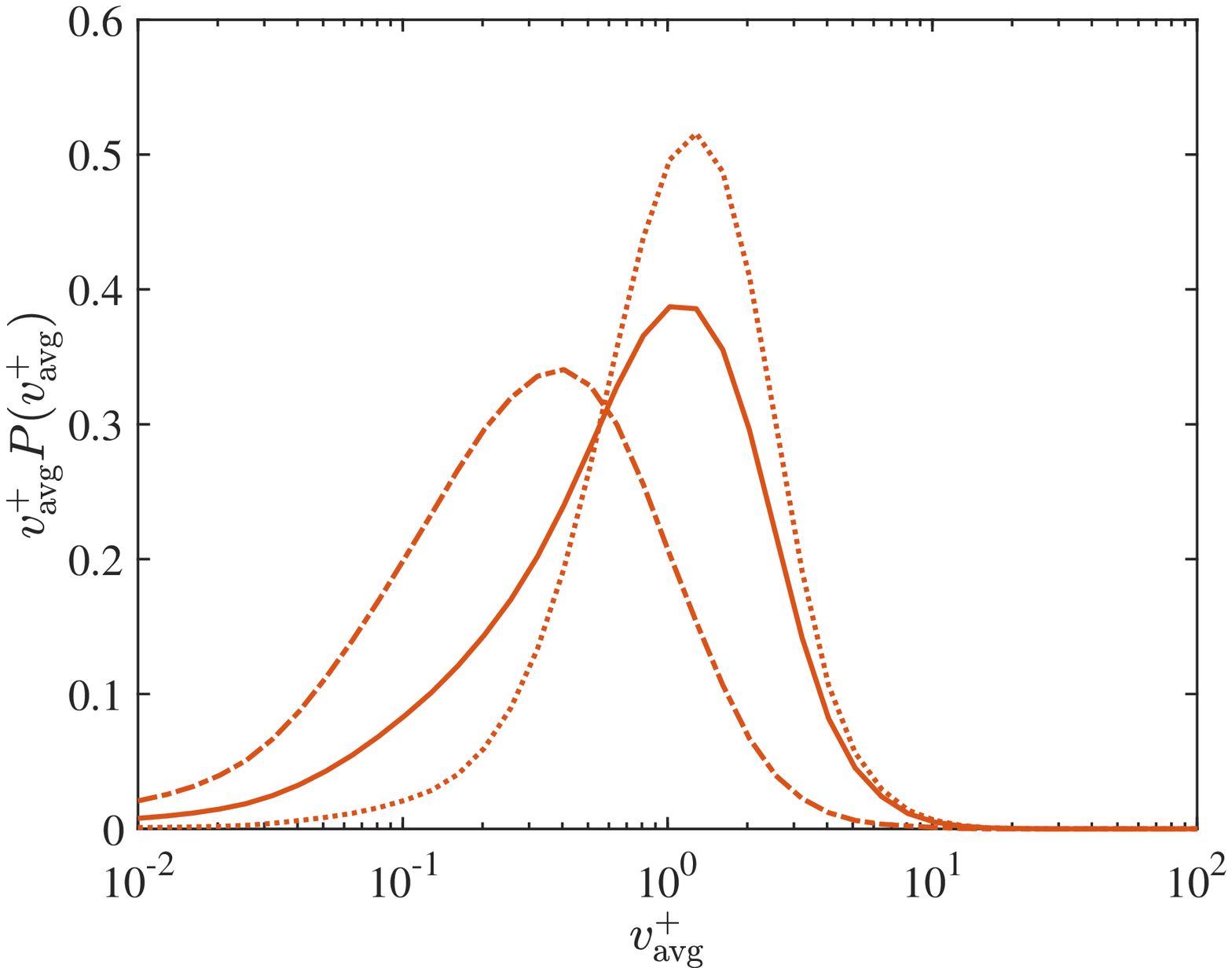}}
\hspace{0.2cm}
\subfloat[]{\includegraphics[height=0.38\textwidth]{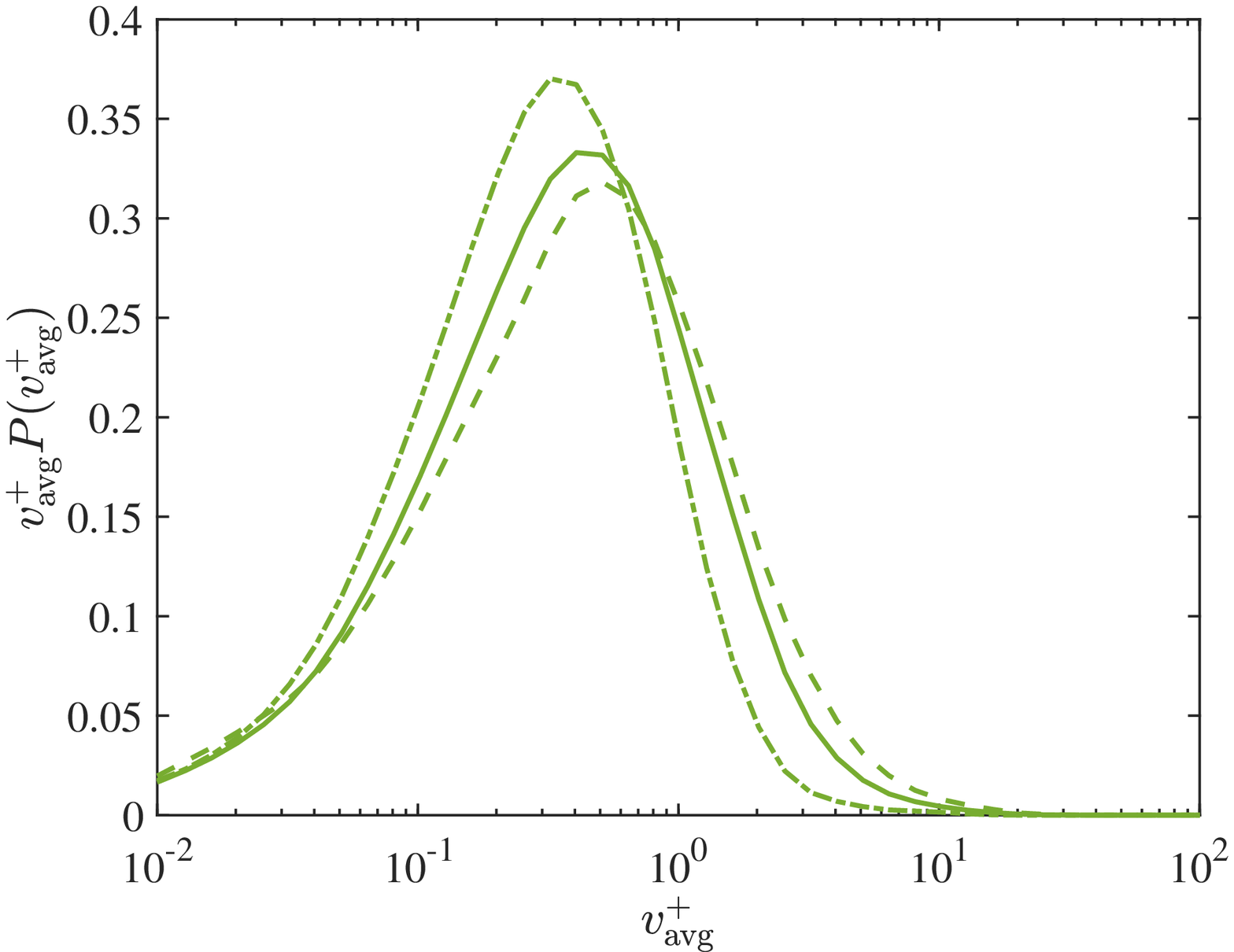}}
\caption{({a}) C.d.f. of the normalized centroid of the streak as a function of the lifetime of a streak. Lines indicate median values of primary (yellow), incoming (blue), outgoing (red), and connector (green) branches. Transparent regions are 25 and 75 percentiles. ({b}) Premultiplied p.d.f. of the average wall-normal velocity of branches for primary (yellow), incoming (blue), outgoing (red), and connector (green) branches. ({c}) Premultiplied p.d.f. of the average wall-normal velocity of outgoing branches (solid), outgoing branches consisting mostly detached streaks (dotted), and outgoing streaks consisting mostly buffer-layer streaks (dot-dashed). ({d}) Premultiplied p.d.f. the average wall-normal velocity of connector branches (solid), connector branches consisting mostly tall-attached streaks (dashed), and outgoing streaks consisting mostly buffer-layer streaks (dot-dashed).}
\label{fig:vavg}
\end{figure}

Finally, we study the relative position of the streak as a function of its lifetime (Figure~\ref{fig:vavg}{a}). All types of streaks move up during their life-cycle -- this is expected as streaks are events with intense $u<0$ events and, thus, are associated with ejections. The wall-normal speed of the streaks tends to be constant throughout their lifetimes, shown by the linear increase in the distance of its centroid from the wall. The distribution of the average wall-normal velocity of these streaks, $v_{\text{avg}}$, for each branch is given in Figure~\ref{fig:vavg}({b}), which shows a similar distribution between the primary, incoming and connector branches, but a positively shifted distribution of wall-normal velocities for the outgoing branches. Conditionally sampling outgoing branches that are composed primarily of detached streaks and those composed primarily of buffer-layer streaks (Figure~\ref{fig:vavg}{c}), we see that the discrepancy is due to the detached streaks, which clearly show a different distribution of wall-normal velocities. The significantly higher positive wall-normal velocities associated with the detached outgoing branches coincide with our observation that detached streaks correlate with bursts or strong ejection events. The incoming and primary branches are mostly composed of buffer layer streaks, so it is difficult to see the p.d.f. breakdown of the average wall-normal velocity conditioned to a particular streak type, but the connector branch is evenly divided among tall-attached and buffer layer streaks. The average wall-normal velocity of the connector branch conditioned to mostly tall-attached streaks and buffer-layer streaks are shown in Figure~\ref{fig:vavg}(d). While there is some discrepancy in the distribution, the difference is not as noticeable as in the case of the detached streaks and similar to the distributions for the incoming and primary branches. This indicates that the detached streaks in the outgoing branch indeed have a significantly stronger wall-normal velocity. This corroborates the theory that these streaks can be identified as bursting events as well as strong Q2 events or ejections.  

\begin{figure}
\centering
{\includegraphics[height=0.42\textwidth]{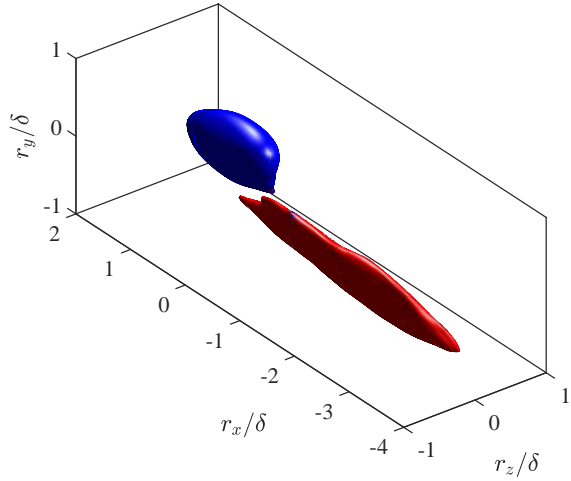}}
\caption {Averaged flow field conditioned to a split event of detached streak from a tall-attached streak. The detached streak is colored in blue for clarity. Isosurfaces are given by $\sqrt{u^2+w^2} > 0.6u_\tau$ and $u < 0$. Direction of the mean streamwise flow is indicated by the arrow.}
\label{fig:cond_avg}
\end{figure}

The averaged flow field conditioned to a split event when a detached streak splits from a tall-attached streak is shown in Figure~\ref{fig:cond_avg}. Conditional averages are computed in the reference frame 
\begin{equation}
    (r_x,r_y,r_z) = \left((x,y,z) - (x_m, y_m, z_m)\right) / \Delta y_m,
\end{equation}
where $(x_m,y_m,z_m)$ is the midpoint of the shortest line connecting the center of the detached streak and the spine of the tall-attached streak and $\Delta y_m$ is the distance from $y_m$ to the wall where the tall-attached streak is attached. The spanwise direction is chosen such that $r_z > 0$ for the tall-attached streak. The center of the conditionally averaged detached streak is located at $r_x\approx0.64, r_y \approx 0.42$, whereas the center of the averaged structure signifying tall-attached streaks is located at $r_x\approx -0.48, r_y \approx -0.64$. This shows that the detached streak splits from the tall-attached streak at a greater (less negative) streamwise and wall-normal velocity, consistent with the observations regarding wall-normal velocities of detached streaks in Figure~\ref{fig:vavg}({c}). The average tall-attached streak shows an elongated shape with aspect ratios similar to the one observed in Figure~\ref{fig:static_jpdf}({a}).

The relative position of the detached streak with respect to the tall-attached streak places the detached streak on the tip of the vortex cluster placed between high-speed and low-speed streaks, similar to the findings regarding Q2 events with respect to conditionally averaged velocity profiles \cite{Johansson1991} and vortex clusters \cite{,delAlamo2006,Lozano-Duran2012}. The relatively higher value of $u$ (less negative) and $v$ (more positive) of the detached streak compared to the tall-attached streak coincides with the direction of the vorticity at the tip of the so-called `horseshoe' or `hairpin' vortices, where concentration of spanwise mean vorticity is accomplished by the creation of large contributions to Reynolds stress. It is worth mentioning that the smooth shape of the conditional averaged quantity is not representative of the individual streaks in the flow, which are more complex and increase in complexity with higher Reynolds numbers. Regardless, the conditional field allows a structural assessment of averaged events related to these events. While the low-speed streaks were argued to be a consequence of the Q2 events in Lozano-Dur\'an \emph{et al.}~\cite{Lozano-Duran2014}, it could also be argued that the flow field conditioned to Q2 events are indeed conditioned to these breakup of streaks and both are consequences of each other.

\section{Conclusions} \label{sec:conclusions}

\begin{figure}
\centering
{\includegraphics[width=0.9\textwidth]{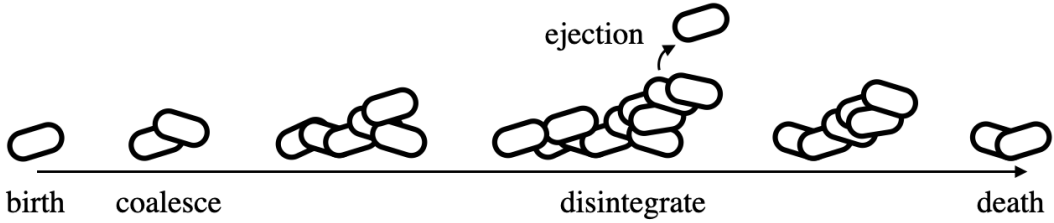}}
\caption {Diagram depicting a life-cycle of a buffer layer streak.}
\label{fig:lifecycle}
\end{figure} 

We study the kinematics and dynamics of buffer layer streaks using temporally resolved DNS data from a turbulent channel at $Re_\tau = 186$. The temporal resolution was fine enough and the duration of the simulation was long enough to track hundreds of thousands of streaks from creation to disintegration. We found that although the interacting streaks could create connected structures that span long spatial domains and time periods, beyond the spatial and temporal domain of the current study, these large structures are composite streaks and can be broken down into much smaller individual streaks that only last a small fraction of that time. 

We first divide the streaks into tall-attached streaks that span from the wall to the outer region, detached streaks that are detached from the wall, and buffer-layer streaks that are wall-attached and stay within the buffer layer. The tall-attached streaks tend to be larger in volume, meander more, but are still aligned with the streamwise direction. The buffer-layer streaks show a similar aspect ratio as the tall-attached streaks but are smaller in size. The detached streaks are more isotropic and tend to be less aligned with the streamwise direction, showing a wide range of azimuth and elevation angles. The distribution of the dimensions of the tall-attached streaks is such that $\Delta x \sim \Delta y^2$ and $\Delta x \sim \Delta z$, and the volume is proportional to $\Delta x^{3/2}$. It was also observed that the meandering is correlated with the $\Delta x/\Delta y$, which increases with volume, indicating that larger streaks start to meander more due to the substantial elongation in the streamwise direction resulting from the larger volume.

The tracking process resulted in the organization of the streaks into branches, which are divided into four categories: primary, incoming, outgoing, and connector. A large portion of the branches identified were connector branches, eluding to the complex merging and splitting of streaks throughout their lifetime. Each branch tends to be composed of streaks that do not change in category, e.g. branches starting as buffer-layer streaks tend to stay buffer-layer streaks during their lifetime. Splitting of branches tends to happen with tall-attached streaks breaking into a tall-attached streak and another streak. Merging events happen most frequently between buffer-layer streaks and tall-attached streaks. The results show that streaks are born in the buffer layer, coalescing with each other to create larger streaks that are still attached to the wall, similar to the observations in a streamwise-minimal channel flow \cite{Toh2005,Abe2018}. Once the streak becomes large enough, the tall-attached streak eventually splits into wall-attached and wall-detached components. The simplified version of the process is summarized in figure \ref{fig:lifecycle}.

The lifetime of the branches depends on the volume of the largest streak within the branch. For incoming and primary branches, this is because the long lifetime allows the streak to grow before it merges into another branch or disintegrated into the turbulent background. For outgoing branches, the initial volume indicates how long the branch will sustain before dissipating. All branches move away from the wall during their lifetime, consistent with the observation that most strong $u<0$ events are Q2 events ($u<0$, $v>0$) \cite{Wallace1977,Wallace2016}. The strongest average wall-normal velocity is observed in outgoing branches primarily composed of detached streaks, which can be seen as strong ejection or bursting events. 

Averaging the velocity fields conditioned to an event where a detached streak splits from a tall-attached streak, we observe that the detached streak splits with a less negative streamwise and a more positive wall-normal velocity compared to the tall-attached streak. The relative position of the detached streak relative to the tall-attached streak is equivalent to the relative positioning of Q2 events with respect to low-speed streak seen in Lozano-Dur\'an \emph{et al.}~\cite{Lozano-Duran2012}. The relatively larger streamwise and wall-normal velocities of the detached streak could also be seen as part of a spanwise vortical structure on top of the vortex cluster resting between the low- and high-speed streaks. The observations allude to the fact that no one structure or event is a cause for all the other events but rather a connected set of events that happen synchronously and can be observed in various ways. Bursting is simultaneously a strong Q2 event as well as attached streaks splitting into detached streaks.

\section*{Acknowledgements}

The authors thank N. Hutchins for his insightful comments on the manuscript.

Sandia National Laboratories is a multimission laboratory managed and operated by National Technology and Engineering Solutions of Sandia, LLC., a wholly owned subsidiary of Honeywell International, Inc., for the U.S. Department of Energy's National Nuclear Security Administration under contract DE-NA0003525. This paper describes objective technical results and analysis. Any subjective views or opinions that might be expressed in the paper do not necessarily represent the views of the U.S. Department of Energy or the United States Government.

\bibliography{streak_id}

\end{document}